\documentclass[10pt,aps,pra,twocolumn,superscriptaddress,floatfix,showpacs,longbibliography]{revtex4-1}
\usepackage{graphicx}
\usepackage{amsfonts}
\usepackage{amssymb}
\usepackage{amsmath}
\usepackage{txfonts}
\usepackage{lipsum}
\usepackage{color}
\usepackage{wasysym}
\usepackage{hyperref}
\usepackage{bbold}
\usepackage{etoolbox} 
\graphicspath{{figures/}}

\usepackage{mathtools}
\usepackage{multirow}
\usepackage{hhline}

\usepackage{mathrsfs} 

\newcommand{\av}[1]{\ensuremath{\left\langle #1 \right\rangle}}
\newcommand{\qv}{\mathbf{q}}
\newcommand{\kv}{\mathbf{k}}

\usepackage{letltxmacro}
\LetLtxMacro{\oldsqrt}{\sqrt}
\renewcommand{\sqrt}[2][\mkern8mu]{\mkern-6mu\mathop{}\oldsqrt[#1]{#2}}

\begin{document}
\title{
Consistent partial bosonization of the extended Hubbard model
}
\author{E. A. Stepanov}
\affiliation{Institute of Theoretical Physics, University of Hamburg, 20355 Hamburg, Germany}
\affiliation{\mbox{Theoretical Physics and Applied Mathematics Department, Ural Federal University, Mira Street 19, 620002 Ekaterinburg, Russia}}

\author{V. Harkov}
\affiliation{Institute of Theoretical Physics, University of Hamburg, 20355 Hamburg, Germany}
\affiliation{European X-Ray Free-Electron Laser Facility, Holzkoppel 4, 22869 Schenefeld, Germany}

\author{A. I. Lichtenstein}
\affiliation{Institute of Theoretical Physics, University of Hamburg, 20355 Hamburg, Germany}
\affiliation{\mbox{Theoretical Physics and Applied Mathematics Department, Ural Federal University, Mira Street 19, 620002 Ekaterinburg, Russia}}
\affiliation{European X-Ray Free-Electron Laser Facility, Holzkoppel 4, 22869 Schenefeld, Germany}

\begin{abstract}
We design an efficient and balanced approach that captures major effects of collective electronic fluctuations in strongly correlated fermionic systems using a simple diagrammatic expansion on a basis of dynamical mean-field theory.
For this aim we perform a partial bosonization of collective fermionic fluctuations in leading channels of instability.
We show that a simultaneous account for different bosonic channels can be done in a consistent way that allows to avoid the famous Fierz ambiguity problem. 
The present method significantly improves a description of an effective screened interaction $W$ in both, charge and spin channels, and has a great potential for application to realistic $GW$-like calculations for magnetic materials. 
\end{abstract}

\maketitle

\section{Introduction}

Mean-field theory is a simple and transparent method that is used for a description of collective fermionic excitations in a broad range of physical problems from condensed matter physics to quantum field theory. It allows to capture both, magnetic and superconducting fluctuations in Hubbard~\cite{Fradkin, STB} and $t$-$J$~\cite{RevModPhys.66.763,PhysRevB.38.5142} models, as well as spontaneous symmetry breaking and formation of various condensates in Nambu--Jona-Lasinio and Gross--Neveu models~\cite{PhysRev.122.345, PhysRev.124.246, BAILIN1984325, ALFORD1998247, BERGES1999215, ALFORD1999443, PhysRevLett.82.3956, PhysRevD.10.3235, ZHUKOVSKY2012597}. The underlying idea of the method is based on a partial bosonization of collective fermionic fluctuations in leading channels of instability in the system~\cite{PhysRevD.68.025020, PhysRevB.70.125111, Jaeckel}. 
This allows a simple diagrammatic solution of the initial problem in terms of original fermionic and effective new bosonic fields in a $GW$ fashion~\cite{GW1, GW2, GW3}.

Theoretical description of many-body effects in a regime of strong electronic interactions requires more advanced approaches that are usually based on (extended) dynamical mean-field theory (EDMFT)~\cite{RevModPhys.68.13, PhysRevB.52.10295, PhysRevLett.77.3391, PhysRevB.61.5184, PhysRevLett.84.3678, PhysRevB.63.115110}. DMFT provides an exact solution of the problem in the limit of infinite dimension~\cite{PhysRevLett.62.324} and is found to be a good approximation for single-particle quantities~\cite{PhysRevB.91.235114}, especially when properties of the system are dominated by local correlations. However, collective electronic fluctuations are essentially nonlocal. For this reason, a number of proposed approaches that treat many-body excitations beyond DMFT grows as fast as a degree of their complexity~\cite{RevModPhys.90.025003}. 
These new methods provide a very accurate solution of model (single-band) problems, but are numerically very expensive for realistic multiband calculations~\cite{PhysRevLett.107.137007, PhysRevB.85.115128, PhysRevB.95.115107, Boehnke_2018, 2018arXiv181105143A, 2019arXiv190407324S}. 

Following the mean-field idea, a partially bosonized description of collective electronic effects in strongly correlated systems can also be performed on a basis of EDMFT. Research in this direction resulted in $GW$+EDMFT~\cite{PhysRevB.66.085120,PhysRevLett.90.086402, PhysRevLett.109.226401, PhysRevB.87.125149, PhysRevB.90.195114, PhysRevB.94.201106, PhysRevB.95.245130} and TRILEX~\cite{PhysRevB.92.115109, PhysRevB.93.235124, PhysRevLett.119.166401} methods.
Although the $GW$-like extension of EDMFT is an efficient and inexpensive numerical approach, it has a significant drawback that is common for every partially bosonized theory.
This severe problem is known as the Fierz ambiguity~\cite{PhysRevD.68.025020, PhysRevB.70.125111, Jaeckel}.
It appears when two or more different bosonic channels are considered simultaneously. Then, the theory becomes drastically dependent on the way how these channels are introduced.
Surprisingly, this issue remains unsolved even for a standard mean-field theory, let alone the $GW$+EDMFT method that is actively used for solution of realistic multiband~\cite{PhysRevLett.90.086402, PhysRevLett.113.266403, Tomczak_2012, PhysRevB.88.165119, PhysRevB.88.235110, PhysRevB.90.165138} and time-dependent problems~\cite{RevModPhys.86.779, PhysRevLett.118.246402}.

Recently, the authors of TRILEX approach showed that the effect of the Fierz ambiguity can be reduced using a cluster extension of the theory~\cite{PhysRevLett.119.166401}. However, this approach is much more time consuming numerically than its original single-site version and, in fact, breaks a translational symmetry of the initial lattice problem. Indeed, the nonlocal in space self-energy obtained within the cluster becomes different from the corresponding one between two clusters. All above discussions suggest that there is no reliable simple theory that can accurately describe an interacting fermionic system in the regime of coexisting strong bosonic fluctuations in different channels.

In this work we introduce a consistent partial bosonization of an extended Hubbard model that solves the famous Fierz ambiguity problem without a complicated cluster extension of the method. 
We show that the resulting action of the problem contains only an effective fermion-boson vertex function, while a fermion-fermion interaction can be safely excluded from the theory. 
The derived approach combines a simplicity of a mean-field approximation with an efficiency of much more advanced EDMFT-based methods. 
This allows to improve many existing extensions of $GW$ method and include an effect of magnetic fluctuations in a standard $GW$ scheme in a consistent way.
Although the introduced theory is discussed in a context of an extended Hubbard model, it is not restricted only to this particular single-band model, and can be applied to other fermionic problems from different areas of physics. 

\section{Partial bosonization of a fermion model}
\subsection{Fierz ambiguity}

We start the derivation of a partially bosonized theory for strongly correlated electrons with the following action of extended Hubbard model
\begin{align}
{\cal S}_{\rm latt} =& -\sum_{\kv,\nu,\sigma} c^{*}_{\kv\nu\sigma} [i\nu+\mu - \varepsilon^{\phantom{*}}_{\kv}] c^{\phantom{*}}_{\kv\nu\sigma} \notag\\
&+ U\sum_{\qv,\omega}n_{\qv\omega\uparrow}n_{-\qv,-\omega\downarrow} + \frac{1}{2}\sum_{\qv,\omega,\varsigma}V^{\varsigma}_{\qv} \, \rho^{\varsigma}_{\qv\omega} \, \rho^{\varsigma}_{-\qv,-\omega}.
\label{eq:actionlatt}
\end{align}
Here, $c^{(*)}_{\kv\nu\sigma}$ is a Grassmann variable corresponding to annihilation (creation) of an electron with momentum $\kv$, fermionic Matsubara frequency $\nu_n$, and spin projection $\sigma=\, \uparrow,\downarrow$. We also introduce following bilinear combinations of fermionic variables $\rho^{\varsigma}_{\qv\omega} = n^{\varsigma}_{\qv\omega} - \av{n^{\varsigma}}$ that correspond to charge ($\varsigma=c$) and spin ($\varsigma=\{x,y,z\}$) degrees of freedom with momentum $\qv$ and bosonic frequency $\omega_m$. $n^{\varsigma}_{\qv\omega} = \sum_{\kv,\nu,\sigma\sigma'}c^{*}_{\kv\nu\sigma}\sigma^{\varsigma}_{\sigma\sigma'} c^{\phantom{*}}_{\kv+\qv,\nu+\omega,\sigma'}$, $\sigma^{c}=\mathbb{1}$, and $\sigma^{x,y,z}$ are Pauli matrices in the spin space.  
$U$ corresponds to a local Coulomb interaction, $V^{\varsigma}_{\qv}$ describes a nonlocal Coulomb and direct exchange interactions in the charge and spin channels, respectively. Dispersion relation $\varepsilon_{\kv}$ can be obtained via a Fourier transform of hopping matrix elements $t_{ij}$ between lattice sites $i$ and $j$. All numerical calculations in this work are performed for a half-filled two-dimensional Hubbard model ($V^{\varsigma}_{\qv}, Y^{\varsigma}_{\omega}=0$) on a square lattice with a nearest-neighbor hopping amplitude $t$. The half of the bandwidth $D=4t=1$ sets the energy scale. The temperature is $T=0.1$.

For a simplified description of many-body effects in the system, leading collective electronic excitations can be partially {\it bosonized}~\cite{PhysRevD.68.025020, PhysRevB.70.125111, Jaeckel}. For this aim, the local interaction term $Un_{\uparrow}n_{\downarrow}$ has to be rewritten in terms of bilinear combinations of fermionic variables as $\frac12\sum_{\varsigma}U^{\varsigma}\rho^{\varsigma}\rho^{\varsigma}$.
This allows to introduce an effective bosonic field for every bilinear combination using the Hubbard--Stratonovich transformation for the total (local and nonlocal) interaction part of the problem~\cite{stratonovich1957method, PhysRevLett.3.77}.
It should be noted, however, that the decoupling of the local Coulomb interaction $U$ into different channels can be done almost arbitrary.
As discussed, for instance, in Ref.~\onlinecite{PhysRevB.92.115109}, 
a free choice for the bare interaction $U^{c}$ in the charge channel immediately fixes the $U^{s}=(U^{c}-U)/3$ value of the spin interaction if all three $s=\{x,y,z\}$ spin channels are introduced simultaneously. The Ising decoupling with $U^{z}=U^{c}-U$ corresponds to the case when only the $z$ component of the spin is considered.
Then, if the initial problem~\eqref{eq:actionlatt} is solved exactly, the result does not depend on the way how the decoupling of $U$ is performed. However, an approximate (mean-field or $GW$-like) solution of the problem dramatically depends on the decoupling~\cite{PhysRevLett.119.166401}. This issue is known as Fierz ambiguity~\cite{PhysRevD.68.025020, PhysRevB.70.125111, Jaeckel}.

\subsection{Collective electronic effects beyond EDMFT}

As follows from the above discussions, the Fierz ambiguity problem can be avoided if the local interaction term $Un_{\uparrow}n_{\downarrow}$ stays undecoupled in its original form. 
However, this form of the interaction prevents any Hubbard--Stratonovich transformation.
Nevertheless, in this case we still can benefit from the idea of (extended) dynamical mean-field theory (EDMFT)~\cite{PhysRevLett.62.324, RevModPhys.68.13, PhysRevB.52.10295, PhysRevLett.77.3391, PhysRevB.61.5184, PhysRevLett.84.3678, PhysRevB.63.115110}, where all local correlations are treated {\it exactly} via an effective local impurity problem 
\begin{align}
\label{eq:actionimp}
{\cal S}^{(i)}_{\text{imp}} =& -\sum_{\nu,\sigma} c^{*}_{\nu\sigma}[i\nu+\mu-\Delta^{\phantom{*}}_{\nu}]c^{\phantom{*}}_{\nu\sigma} \notag\\
&+ U\sum_{\omega}n_{\omega\uparrow}n_{-\omega\downarrow} + \frac{1}{2}\sum_{\omega, \varsigma} Y^{\varsigma}_{\omega} \, \rho^{\varsigma}_{\omega} \, \rho^{\varsigma}_{-\omega}.
\end{align}
The latter is a local part of the lattice action~\eqref{eq:actionlatt}, where a dispersion relation and nonlocal interaction are replaced by local fermionic ($\varepsilon_{\kv}\to\Delta_{\nu}$) and bosonic ($V^{\varsigma}_{\qv}\to Y^{\varsigma}_{\omega}$) hybridization functions that effectively account for nonlocal single- and two-particle fluctuations, respectively. 
In the absence of these hybridizations, EDMFT reduces to a static mean-field approximation. Since the impurity model is solved numerically exactly using, e.g., continuous-time quantum  Monte Carlo solvers~\cite{PhysRevB.72.035122, PhysRevLett.97.076405, PhysRevLett.104.146401, RevModPhys.83.349}, the Fierz ambiguity problem on the local level is absent by construction.

Further, we integrate out the impurity problem in order to exactly account for all local fluctuations in the effective lattice model. 
As shown in the dual fermion (DF) approach~\cite{PhysRevB.77.033101}, this can be done after the nonlocal part of the lattice action is rewritten in terms of new fermionic variables $c^{(*)} \to f^{(*)}$. In addition, we perform a partial bosonization $\rho^{\varsigma}\to\varphi^{\varsigma}$ of the nonlocal interaction following the dual boson (DB) scheme~\cite{Rubtsov20121320, PhysRevB.93.045107}, which does not lead to the Fierz ambiguity either. 
Then, the initial problem~\eqref{eq:actionlatt} transforms to a dual action (see Ref.~\onlinecite{PhysRevB.94.205110} and Appendix~\ref{App:Action})
\begin{align}
\label{eq:dualaction}
{\cal \tilde{S}}
= -\sum_{\kv,\nu,\sigma} f^{*}_{\kv\nu\sigma}\tilde{\cal G}^{-1}_{\kv\nu\sigma}f^{\phantom{*}}_{\kv\nu\sigma} 
-\frac12\sum_{\qv,\omega,\varsigma}  \varphi^{\varsigma}_{\qv\omega}
\tilde{\cal W}^{\varsigma~-1}_{\qv\omega}
 \varphi^{\varsigma}_{-\qv,-\omega} + \tilde{\cal F}.
\end{align}
After the impurity problem is integrated out, bare fermion $\tilde{\cal G}_{\kv\nu\sigma}=G^{\rm EDMFT}_{\kv\nu\sigma} - g^{\phantom{E}}_{\nu\sigma}$ and boson $\tilde{\cal W}^{\varsigma}_{\qv\omega} = W^{\rm \varsigma\,EDMFT}_{\qv\omega} - w^{\varsigma}_{\omega}$ propagators 
are given by nonlocal parts of EDMFT Green's function and renormalized interaction~\cite{PhysRevB.94.205110}, respectively. 
Thus, they already account for local single- and two-particle fluctuations in the system via an exact local self-energy $\Sigma^{\rm imp}_{\nu\sigma}$ and polarization operator $\Pi^{\varsigma \, \rm imp}_{\omega}$ of the effective impurity problem, respectively.
Here, $g_{\nu\sigma}$ and $w^{\varsigma}_{\omega}$ are the full local Green's function and renormalized interaction of the impurity problem.  

The interaction part $\tilde{\cal F}[f,\varphi]$ of the dual action~\eqref{eq:dualaction} contains all possible fully screened local fermion-fermion and fermion-boson vertex functions of the impurity problem~\cite{Rubtsov20121320, PhysRevB.93.045107}. Here, as well as in most of DB approximations, we restrict ourselves to the lowest-order (two-particle) interaction terms that are given by the fermion-fermion $\Gamma_{\nu\nu'\omega}$ and fermion-boson $\Lambda_{\nu\omega}$ vertex functions
\begin{align}
\Gamma_{\nu\nu'\omega} = \vcenter{\hbox{\includegraphics[width=0.12\linewidth]{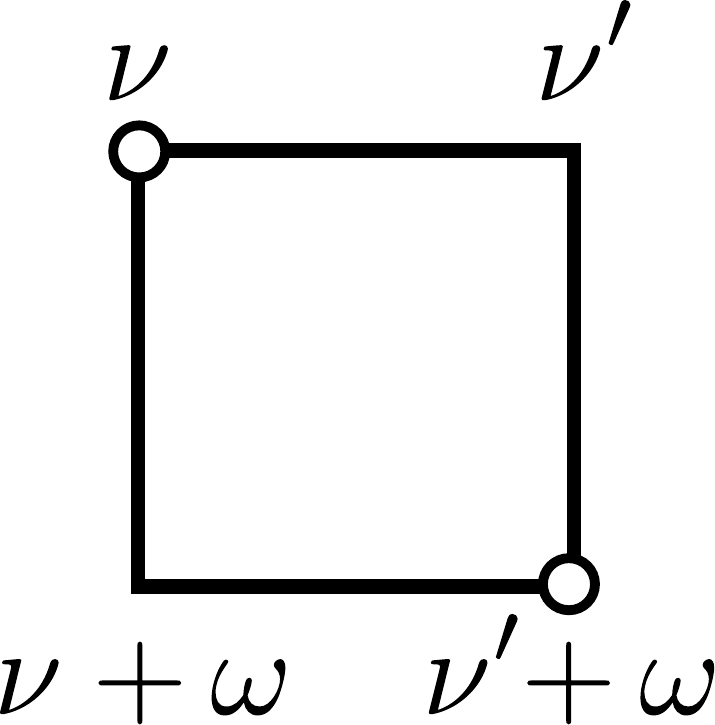}}}, ~~~
\Lambda_{\nu\omega} = \vcenter{\hbox{\includegraphics[width=0.12\linewidth]{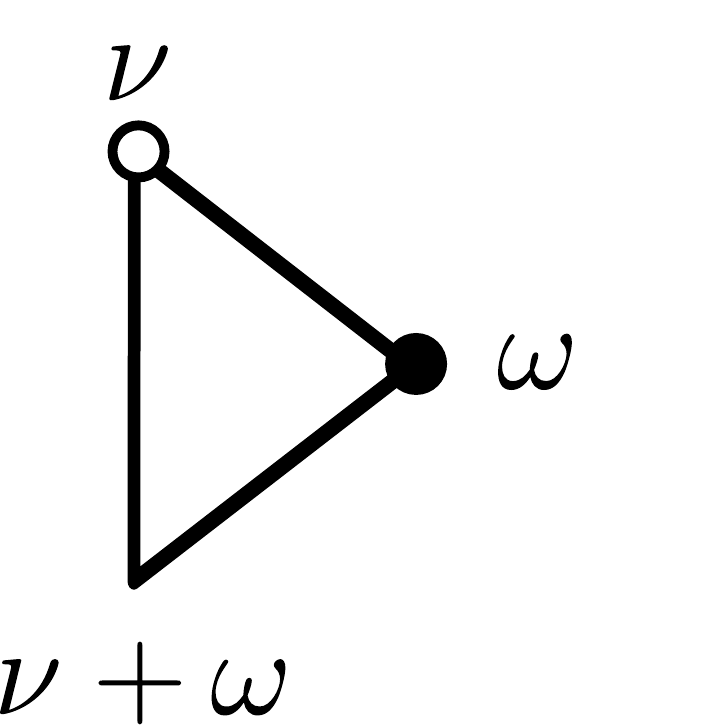}}}.
\end{align}
Exact definition of these quantities can be found in Appendix~\ref{App:Action}. 
The dual theory with only two-particle interaction terms has been tested against exact benchmark results showing a good performance of the theory in a broad regime of model parameters~\cite{PhysRevB.96.035152, PhysRevB.94.035102, PhysRevB.97.125114}. Moreover, the fact that the screened six-fermion vertex function has only a minor effect on the self-energy of the Hubbard model has been observed in~\cite{PhysRevLett.102.206401}.

In the absence of the interaction part $\tilde{\cal F}[f,\varphi]$ the dual theory~\eqref{eq:dualaction} reduces to
EDMFT~\cite{Rubtsov20121320, PhysRevB.93.045107}. However, an account for vertex corrections beyond the dynamical mean-field solution is desirable~\cite{PhysRevMaterials.1.043803, PhysRevLett.109.237010, 2019arXiv190611853Z}.
Especially, it is an important problem for description of spin fluctuations and magnetic polarization in realistic systems~\cite{PhysRevLett.100.116402, PhysRevB.95.041112} as they are not captured by a standard $GW$+DMFT scheme~\cite{PhysRevB.66.085120,PhysRevLett.90.086402, PhysRevLett.109.226401, PhysRevB.87.125149, PhysRevB.90.195114, PhysRevB.94.201106, PhysRevB.95.245130}. While the use of the fermion-boson vertex in a diagrammatic solution of multiband problems is possible~\cite{PhysRevLett.119.166401}, an inclusion of the fermion-fermion vertex in realistic calculations is extremely challenging and time consuming numerically~\cite{PhysRevLett.107.137007, PhysRevB.85.115128, PhysRevB.95.115107, Boehnke_2018, 2018arXiv181105143A, 2019arXiv190407324S}. 
The fermion-fermion vertex describes the full (renormalized) local fermion-fermion interaction, so it cannot be simply discarded. 

It would be extremely helpful to find an additional transformation of the problem~\eqref{eq:dualaction} in which the full local fermion-fermion vertex function $\Gamma_{\nu\nu'\omega}$ vanishes from the effective action.
Then, the resulting theory will be written in terms of fermion and boson propagators, and the remaining fermion-boson interaction $\Lambda_{\nu\omega}$. An effective fermion-fermion vertex function in this theory appears only after bosonic fields are integrated out. Such a fermion-fermion vertex is by definition reducible with respect to a bosonic propagator and serves as an approximation for the original fermion-fermion vertex function $\Gamma_{\nu\nu'\omega}$. 
Since irreducible contributions are not contained in this approximation, the effective fermion-fermion vertex of the resulting fermion-boson theory becomes drastically dependent on the way how bosonic fields are introduced. This fact again leads to the Fierz ambiguity problem. 

\subsection{Approximation for the fermion-fermion vertex}

We have found a unique form of the bare interaction in every considered bosonic channel that almost fully suppresses the effect of missing irreducible diagrams. 
As a consequence, an effective reducible fermion-fermion interaction almost exactly coincides with the full local fermion-fermion vertex  $\Gamma_{\nu\nu'\omega}$, which automatically solves the Fierz ambiguity problem. This unique form of the bare interaction can be found by analyzing the bare fermion-fermion vertex of the impurity problem. 
Let us arbitrarily decouple the local Coulomb interaction $U$ of the impurity problem~\eqref{eq:actionimp} into charge $U^{c}$ and spin $U^{s}$ parts. This leads to the following bare interaction ${\cal U}^{\varsigma}_{\omega}=U^{\varsigma} + Y^{\varsigma}_{\omega}$ in a corresponding bosonic channel. Then, we rewrite the interaction part of the impurity problem in an antisymmetrized form of the bare 
fermion-fermion vertex $\Gamma^{\,0}_{\nu\nu'\omega}$
\begin{align}
&{\cal S}^{(i)}_{\text{imp}} = -\sum_{\nu,\sigma} c^{*}_{\nu\sigma}[i\nu+\mu-\Delta^{\phantom{*}}_{\nu}]c^{\phantom{*}}_{\nu\sigma} \notag\\
&+ \frac{1}{8} \sum_{\nu,\nu',\omega}\sum_{\varsigma,\sigma(')} \Gamma^{\,0\,\varsigma}_{\nu\nu'\omega} c^{*}_{\nu\sigma}\sigma^{\varsigma}_{\sigma\sigma'} c^{\phantom{*}}_{\nu+\omega,\sigma'} c^{*}_{\nu'+\omega,\sigma''}\sigma^{\varsigma}_{\sigma''\sigma'''} c^{\phantom{*}}_{\nu',\sigma'''}.
\label{eq:action_imp_G}
\end{align}
This procedure can be performed in a standard way (see, for instance, Section II A in Ref.~\onlinecite{PhysRevB.57.6884}) interchanging indices of two creation (or annihilation) Grassmann variables in the interaction term. Charge and spin ``$z$'' components of the bare fermion-fermion vertex are given by the expressions
\begin{align}
\Gamma^{\,0\,c}_{\nu\nu'\omega} &= 2{\cal U}^{c}_{\omega} - {\cal U}^{c}_{\nu'-\nu} - {\cal U}^{x}_{\nu'-\nu} - {\cal U}^{y}_{\nu'-\nu} - {\cal U}^{z}_{\nu'-\nu} \notag\\
&= U + 2Y^{c}_{\omega} - Y^{c}_{\nu'-\nu} - Y^{x}_{\nu'-\nu} - Y^{y}_{\nu'-\nu} - Y^{z}_{\nu'-\nu}, \notag\\
\Gamma^{\,0\,z}_{\nu\nu'\omega} &= 2{\cal U}^{z}_{\omega} - {\cal U}^{z}_{\nu'-\nu} + {\cal U}^{x}_{\nu'-\nu} + {\cal U}^{y}_{\nu'-\nu} - {\cal U}^{c}_{\nu'-\nu} \notag\\
&= -U + 2Y^{z}_{\omega} - Y^{z}_{\nu'-\nu} + Y^{x}_{\nu'-\nu} + Y^{y}_{\nu'-\nu} - Y^{c}_{\nu'-\nu}, \label{eq:BareGammaSpin}
\end{align} 
and spin ``$x$'' and ``$y$'' components can be obtained by a circle permutation of spin $\{x,y,z\}$ indices in the second equation. 

\begin{figure}[t!]
\includegraphics[width=0.9\linewidth]{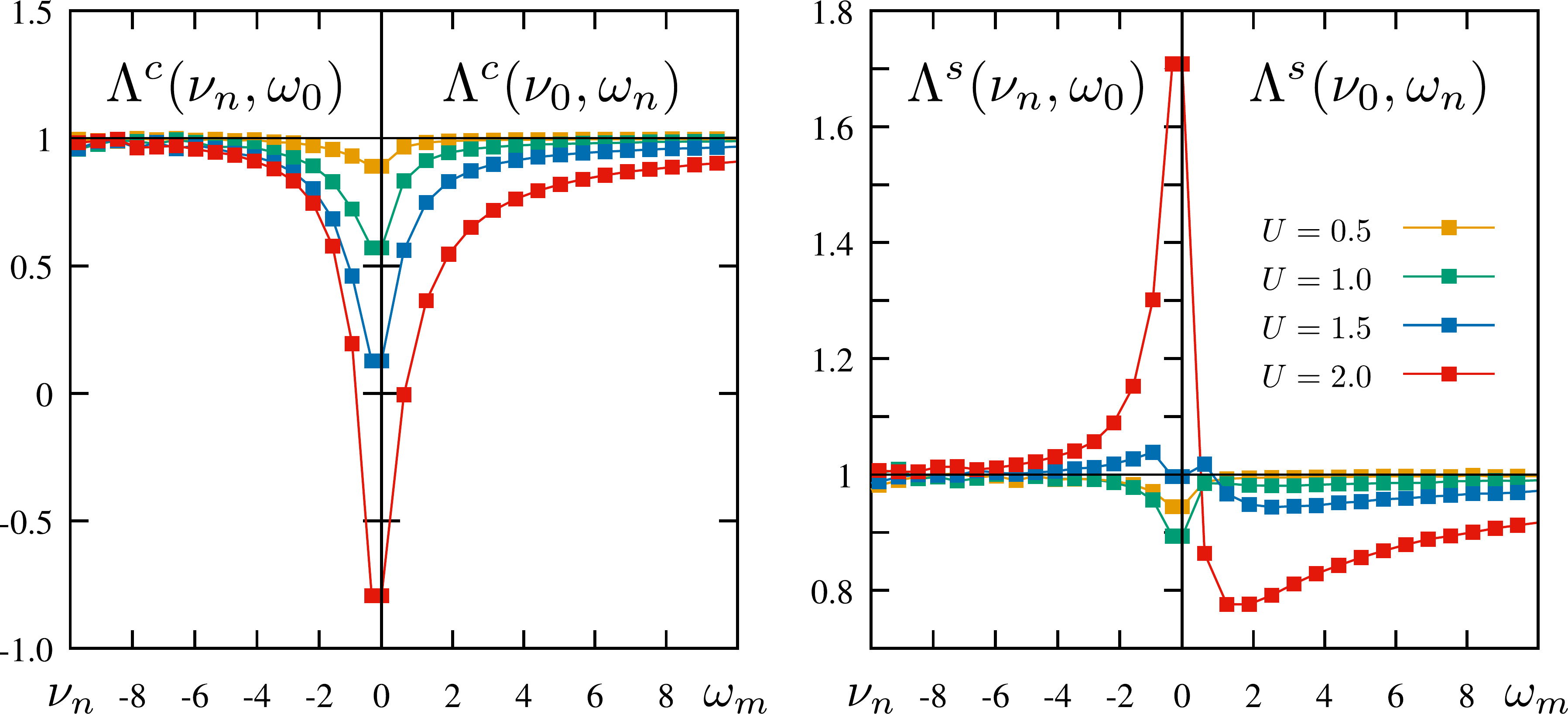}
\caption{Fermion-boson vertex function $\Lambda_{\nu\omega}$ in the charge (left) and spin (right) channels as a function of fermionic $\nu_n$ and bosonic $\omega_{m}$ frequencies. The result is obtained for different values of the local Coulomb interaction.}
\label{fig:Lambda}
\end{figure}

As can be seen from Eq.~\eqref{eq:BareGammaSpin}, the ladder-like irreducible contributions to the fermion-fermion vertex $\Gamma^{\,\varsigma}_{\nu\nu'\omega}$ of the impurity problem originate from the presence of ``vertical'' bosonic lines ${\cal U}^{\varsigma}_{\nu'-\nu}$ in the bare vertex.
Dressed by a two-particle ladder they become irreducible with respect to the (``horizontal'') bosonic line ${\cal U}^{\varsigma}_{\omega}$ and will not be included in the reducible approximation.
As the second line in Eq.~\eqref{eq:BareGammaSpin} shows, the bare vertex $\Gamma^{\,0\,\varsigma}_{\nu\nu'\omega}$ does not depend on the way how the decoupling of the local Coulomb interaction is performed. This fact follows from the exact relation between bare interactions $U^{\varsigma}$ in different bosonic channels.
Therefore, let us include the main contribution $\pm{}U$ of the charge/spin bare vertex only to the horizontal line ${\cal U}^{\varsigma}_{\omega}$.
This immediately leads to a unique form of the bare interaction $U^{c} = -U^{s} = U/2$ with the same value for all $s=\{x, y, z\}$ spin components that excludes ladder-like irreducible contributions from the full local fermion-fermion vertex function. If more complicated non-ladder irreducible contributions to the fermion-fermion vertex become important, they cannot be completely excluded from the theory, but are still strongly suppressed by our choice of the bare interaction. Importantly, this result for the bare interaction {\it cannot} be obtained by any decoupling of the Coulomb interaction $U$ discussed above. 
Note that the fermion-boson vertex is by definition irreducible with respect to the bosonic propagator, the inclusion of the full local Coulomb interaction $U$ in the horizontal line leads to a correct asymptotic behavior of this vertex $\Lambda^{c/s}_{\nu\omega}\to1$ at large frequencies as shown in Fig.~\ref{fig:Lambda}.

The best possible decoupling-based approximation for the fermion-fermion vertex can be obtained for the Ising form of the bare interaction $U^{c} = -U^{z} = U/2$ and $U^{x},U^{y}=0$.
This approximation still reproduces the ``$-U$'' contribution to the bare vertex $\Gamma^{\,0\,x/y}_{\nu\nu'\omega}$ via $U^{c}$ and $U^{z}$ terms, but neglects the screening of this vertex by two-particle fluctuations in $x$ and $y$ channels. Note that the Ising decoupling leads to a correct Hartree-Fock saddle point in the mean-field description of spin fluctuations~\cite{PhysRevLett.65.2462}. Moreover, the Ising decoupling provides the best possible result for a single-site TRILEX approach~\cite{PhysRevLett.119.166401}. However, as we show below, the result for physical observables, such as the self-energy, can be drastically improved using our unique form of the bare interaction, which is not based on the decoupling ideology.

\begin{figure}[t!]
\includegraphics[width=0.95\linewidth]{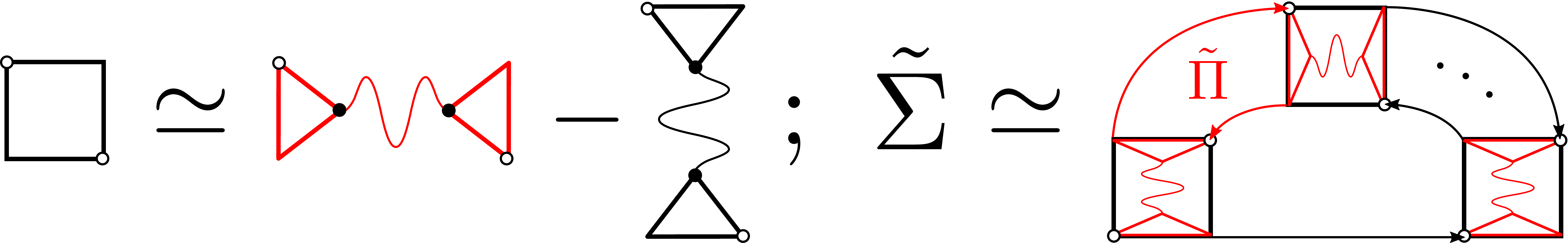}
\caption{The sketch of the approximation for the full local fermion-fermion vertex function $\Gamma_{\nu\nu'\omega}$ introduced in Eq.~\eqref{eq:SpinGamma} (left). The illustration of the TRILEX$^2$ approximation of the nonlocal self-energy $\tilde{\Sigma}$ of the ladder dual theory that accounts only for the horizontal (shown in red) contribution to the fermion-fermion vertex function (right).}
\label{fig:GammaSigma}
\end{figure}

After all, a final result for the reducible approximation of the full fermion-fermion vertex function of the impurity problem can be written in the following form (see Fig.~\ref{fig:GammaSigma})
\begin{align}
\Gamma^{c}_{\nu\nu'\omega} &= 2M^{c}_{\nu\nu'\omega} - M^{c}_{\nu,\nu+\omega,\nu'-\nu} - 3M^{s}_{\nu,\nu+\omega,\nu'-\nu}, \notag\\
\Gamma^{\,s}_{\nu\nu'\omega} &= 2M^{s}_{\nu\nu'\omega} + M^{s}_{\nu,\nu+\omega,\nu'-\nu} - M^{c}_{\nu,\nu+\omega,\nu'-\nu}, \label{eq:SpinGamma}
\end{align}
where 
\begin{align}
M^{\varsigma}_{\nu\nu'\omega} = \Lambda^{\varsigma}_{\nu\omega} w^{\varsigma}_{\omega} \, \Lambda^{\varsigma}_{\nu'+\omega,-\omega} - U^{\varsigma}/2 = \vcenter{\hbox{\includegraphics[width=0.3\linewidth]{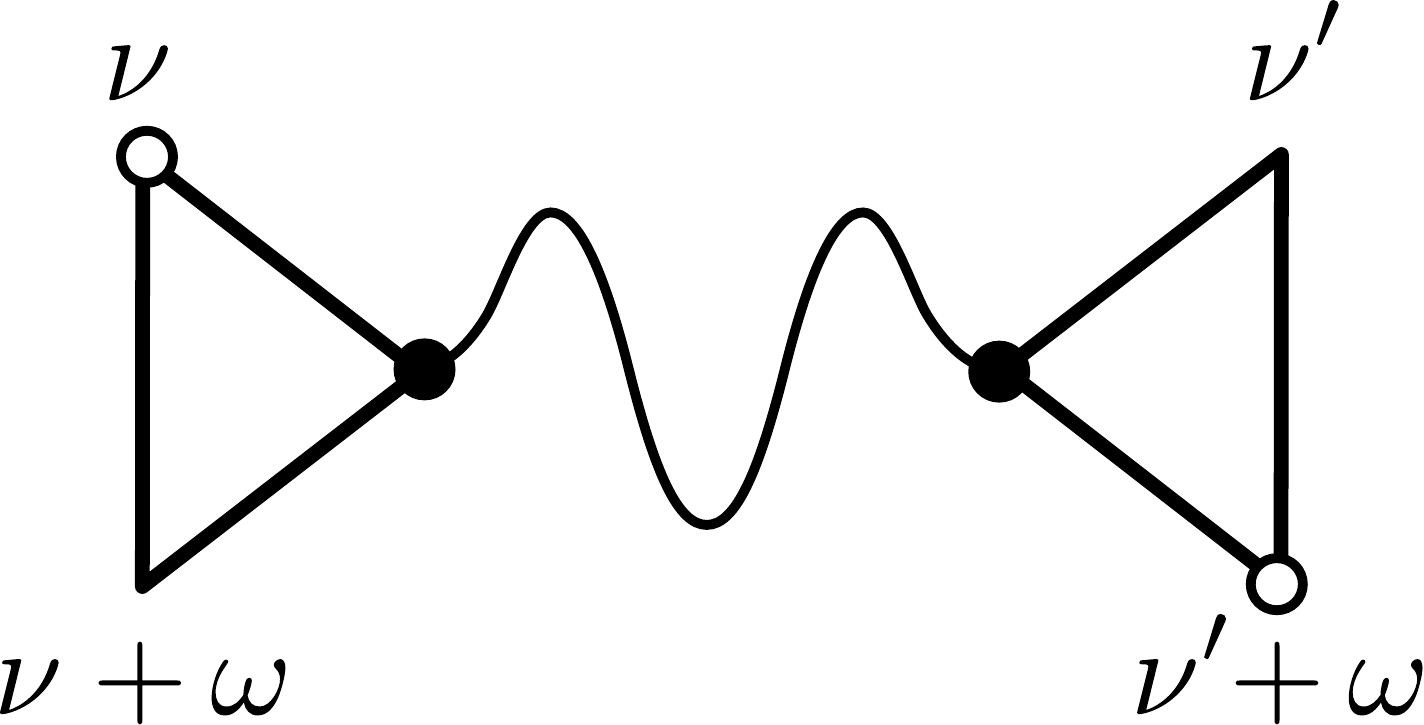}}}~,
\label{eq:correction}
\end{align}
and the term $U^{\varsigma}/2$ excludes a double-counting of the bare Coulomb interaction between different channels. Note that in the case of Ising decoupling of the Coulomb interaction the term $U^{\varsigma}/2$ does not appear in Eq.~\eqref{eq:correction}, because this form of the decoupling is identical for every bosonic channel and does not lead to a double counting.
A detailed derivation of these expressions can be found in Appendix~\ref{App:Vertex}.

A simpler parametrization of the fermion-fermion vertex, which is based on a weak coupling perturbation expansion, has been derived in Refs.~\onlinecite{PhysRevB.79.195125, PhysRevB.81.235108, PhysRevLett.114.236402}. 
A more advanced approximation that additionally accounts for fermion-boson vertex corrections $\Lambda^{\varsigma}_{\nu,\omega}$ has been later introduced in~\cite{PhysRevLett.121.037204, PhysRevB.99.115124}. There, a decomposition of the local Coulomb interaction in only one (spin or charge) channel has been considered.
Note also that in these two works the approximation for the fermion-fermion vertex appears in a nonsymmetrized form that contains only a horizontal contribution $M^{\varsigma}_{\nu\nu'\omega}$~\eqref{eq:correction}. However, it can be {\it identically} rewritten in the antisymmetrized form of Eq.~\eqref{eq:SpinGamma} that has both, horizontal $M^{\varsigma}_{\nu\nu'\omega}$ {\it and} vertical $M^{\varsigma}_{\nu,\nu+\omega,\nu'-\nu}$ components.

\begin{figure}
\includegraphics[width=0.75\linewidth]{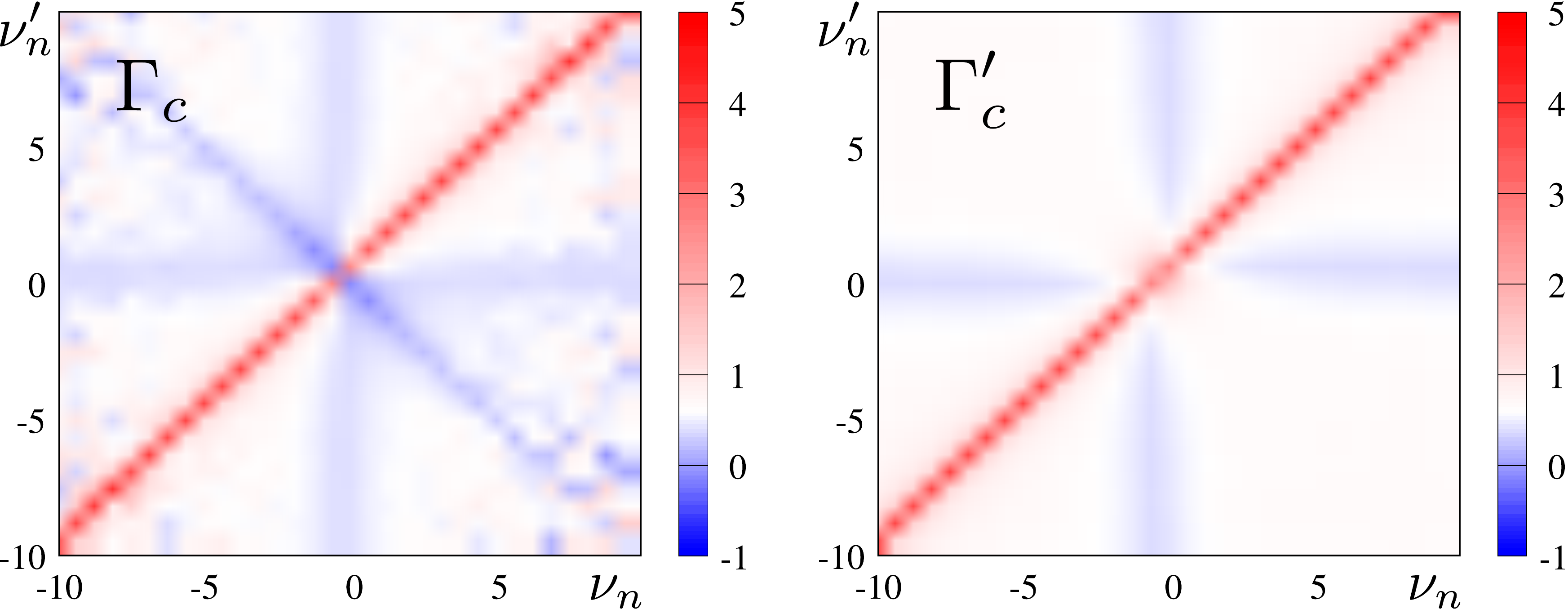} \hspace{1.5cm}
\includegraphics[width=0.75\linewidth]{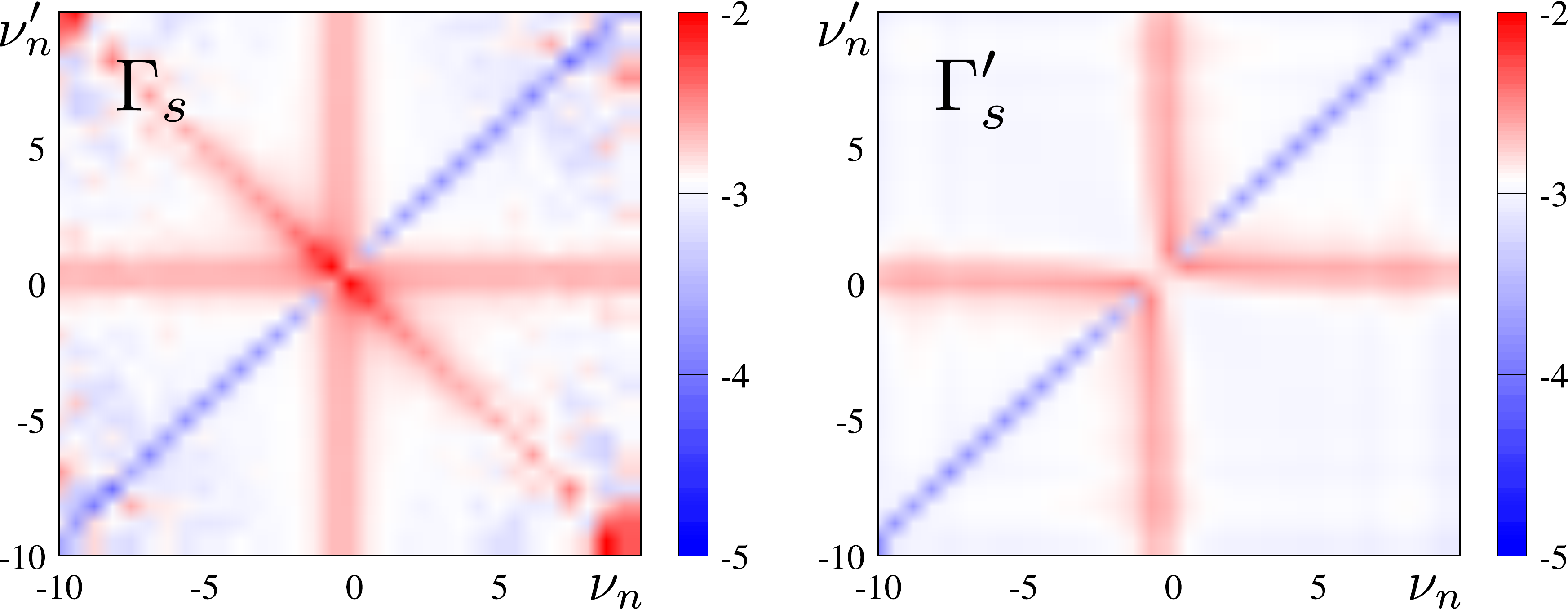}
\caption{Charge and spin components of the exact ($\Gamma_{\nu,\nu'\omega}$) and approximate ($\Gamma'_{\nu,\nu'\omega}$) fermion-fermion vertex functions at zeroth bosonic frequency $\omega_0$. The result is obtained for $U=1.0$.}
\label{fig:Gchsp}
\end{figure}

Our present parametrization~\eqref{eq:SpinGamma} improves the idea of Refs.~\onlinecite{PhysRevLett.121.037204, PhysRevB.99.115124} and exploits a unique multiple channel decomposition of the fermion-fermion vertex. 
We find that this approximation~\eqref{eq:SpinGamma} is in a good agreement with the exact result not only in the weakly interacting regime $U=0.5$, but also at much larger values of the local Coulomb interaction $U=1.0$ and $1.5$.
For this reason, Fig.~\ref{fig:Gchsp} shows the result for the exact and approximate vertex functions only for $U=1.0$, which were obtained for the same impurity problem of dynamical mean-field theory.
Note that the contribution from the particle-particle channel, which at $\omega_0$ is located along the $\nu_n=-\nu'_n$ line~\cite{PhysRevB.86.125114,2016arXiv161006520W}, is not considered in our approximation. Although this contribution to the fermion-fermion vertex is not small itself, it has a minor effect on physical observables, such as a self-energy, at general fillings~\cite{PhysRevB.49.1586}. The exclusion of a particle-particle channel from the approximation of the vertex greatly simplifies the theory as it does not require the calculation of the ``anomalous'' fermion-boson vertex function with two incoming or two outgoing fermionic lines. However, if a certain physical problem needs an account for the particle-particle channel, the latter can be introduced in the theory in the same way as it is done for the particle-hole (charge and spin) channel.
We have noticed that a similar decomposition of the fermion-fermion vertex is proposed in~\cite{2019arXiv190703581K}. 
In contrast, our derivation of an approximate fermion-fermion vertex aims to explain why irreducible contributions are almost fully suppressed by the unique choice of the bare interaction. 
This is a key ingredient for our study that allows to exclude the fermion-fermion vertex function from the theory.

\begin{figure}[t!]
\includegraphics[width=0.9\linewidth]{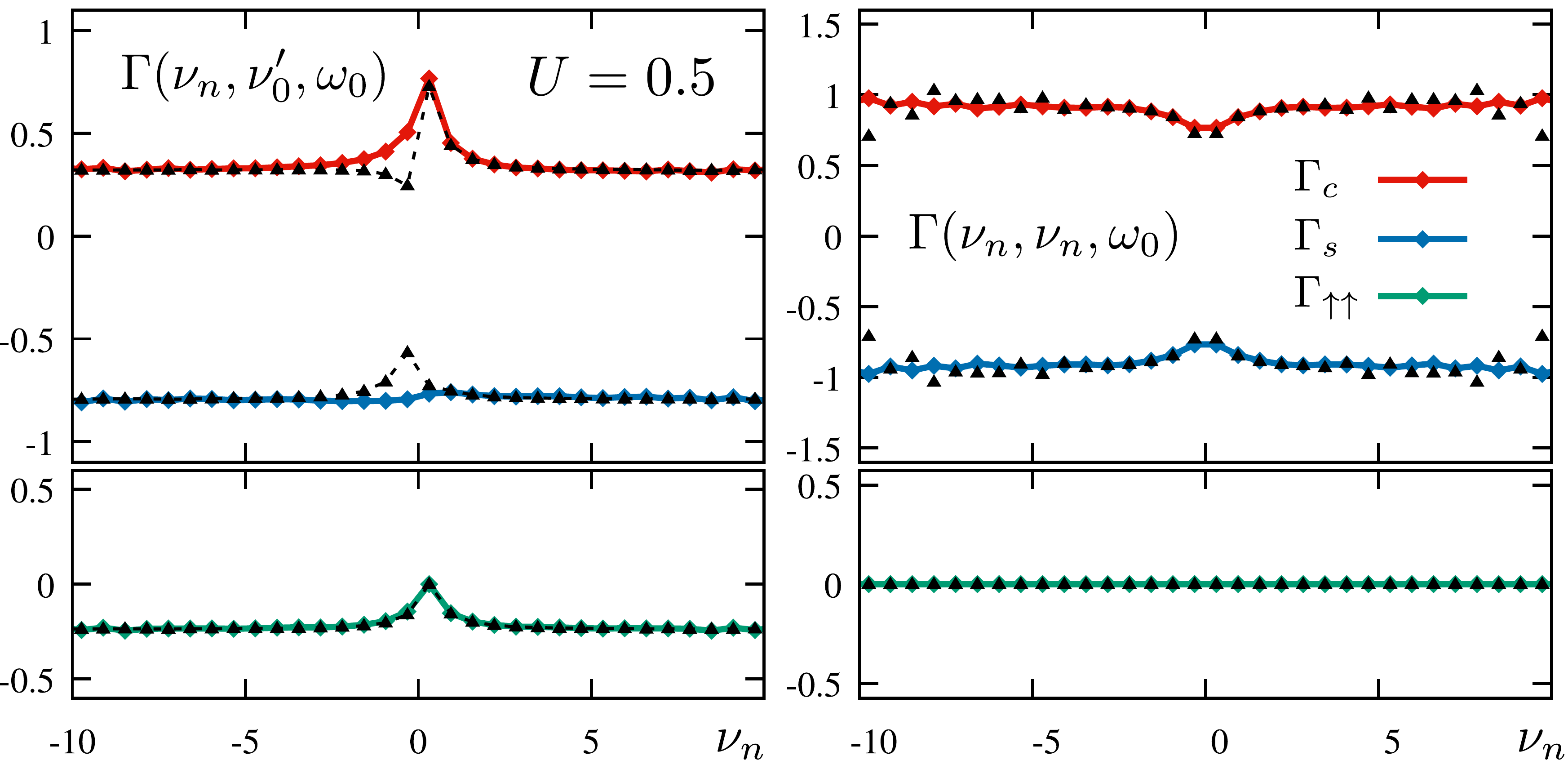} \\[0.3cm]
\includegraphics[width=0.9\linewidth]{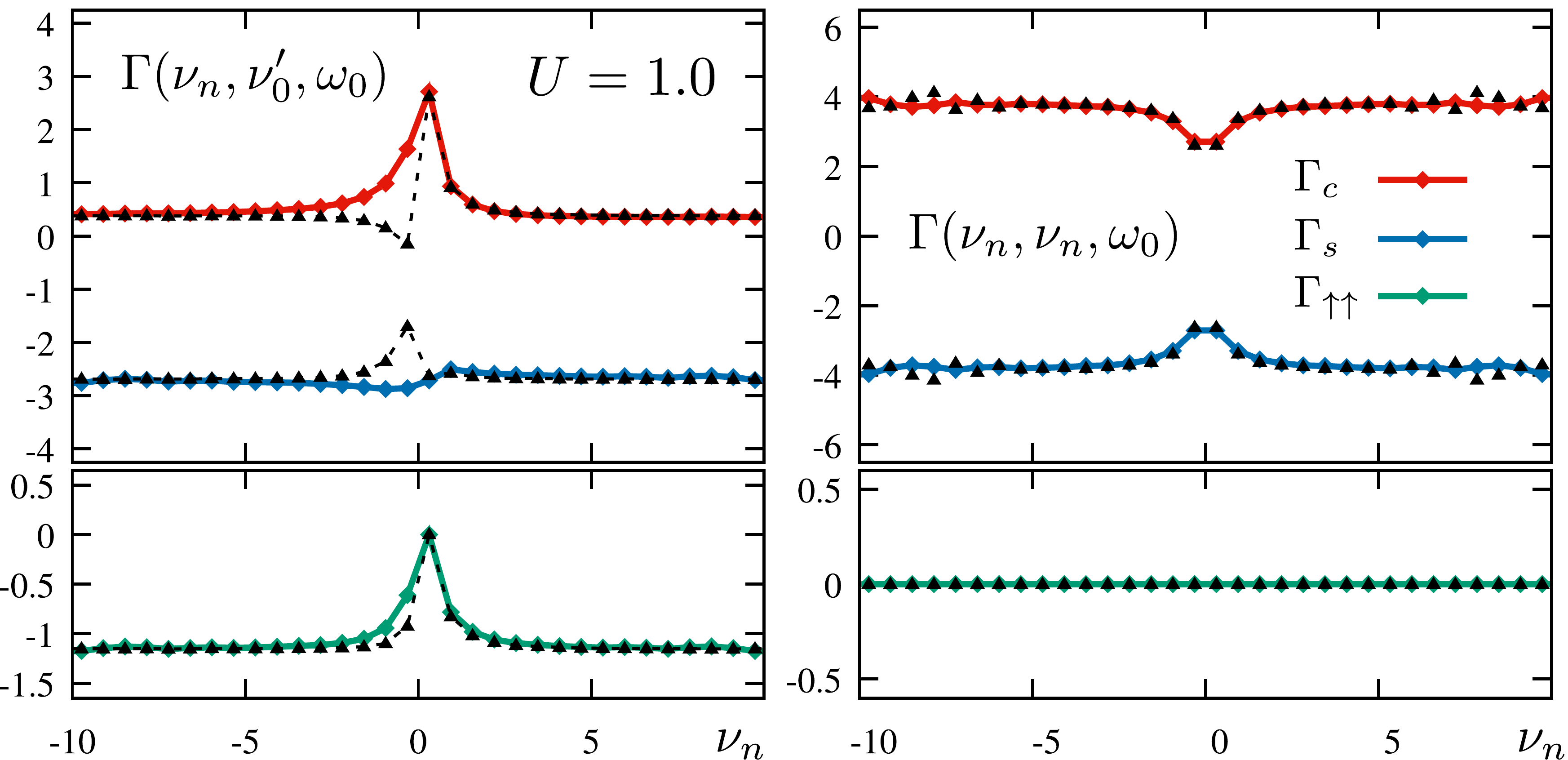} \\[0.3cm]
\includegraphics[width=0.9\linewidth]{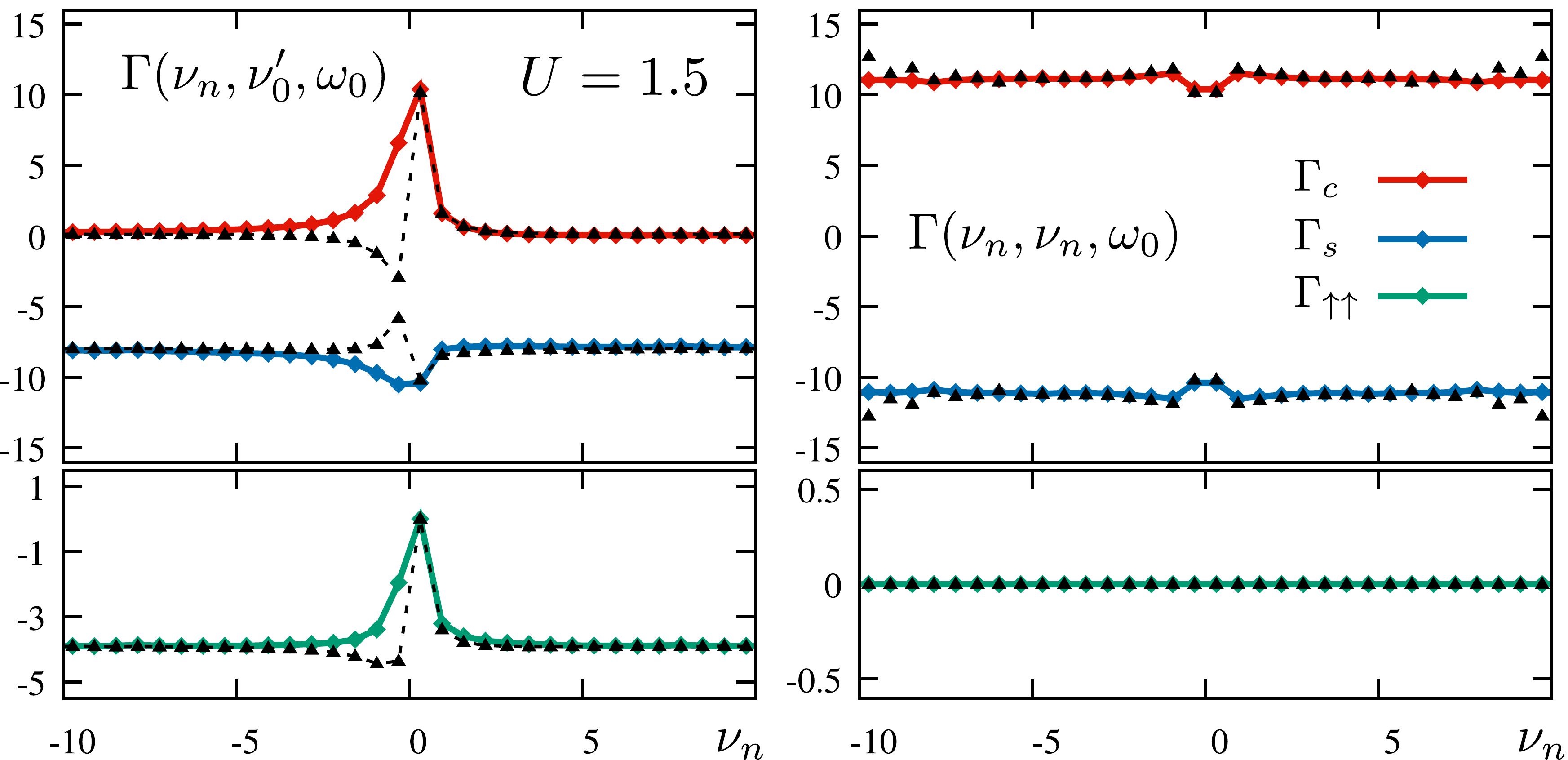} 
\caption{Frequency dependence of charge, spin and $\uparrow\uparrow$ components of the exact (black triangles) and approximate (lines with diamonds) fermion-fermion vertex function $\Gamma_{\nu,\nu'\omega}$ along $\nu'_0$ (left column) and $\nu_n=\nu'_n$ (right column) lines at zeroth bosonic frequency $\omega_0$. Results are obtained for $U=0.5$ (top row), $U=1.0$ (middle row) and $U=1.5$ (bottom row).}
\label{fig:Gamma}
\end{figure}

Figure~\ref{fig:Gamma} shows the cut of the fermion-fermion vertex function $\Gamma_{\nu,\nu',\omega}$ obtained for $U=0.5$ (top row), $U=1.0$ (middle row), and $U=1.5$ (bottom row) at zeroth bosonic frequency $\omega_0$ in two most important directions. We find that the frequency dependence of the exact vertex along $\nu'_{0}$ (left column) and $\nu_{n}=\nu'_{n}$ (right column) lines is captured reasonably well by the horizontal $M^{\varsigma}_{\nu\nu'\omega}$ and vertical $M^{\varsigma}_{\nu,\nu+\omega,\nu'-\nu}$ diagrams, respectively.
A neglected particle-particle contribution results in a mismatch between the approximate and exact results for the fermion-fermion vertex in a small region around the $\nu_{-1}$ point. Since the particle-particle contribution has a minor effect on the $\uparrow\uparrow$ component of the vertex~\cite{PhysRevLett.114.236402}, our approximation provides a reasonably good result for $\Gamma^{\uparrow\uparrow}_{\nu\nu'\omega}=(\Gamma^{c}_{\nu,\nu',\omega}+\Gamma^{s}_{\nu,\nu',\omega})/2$. 

\subsection{Effective fermion-boson model}

Further, we make an additional approximation for the reducible fermion-fermion vertex $M^{\varsigma}_{\nu\nu'\omega}\simeq\Lambda^{\varsigma}_{\nu\omega} \bar{w}^{\varsigma}_{\omega} \, \Lambda^{\varsigma}_{\nu'+\omega,-\omega}$ including the term $U^{\varsigma}/2$ in the propagator $\bar{w}^{\varsigma}_{\omega} = w^{\varsigma}_{\omega} - U^{\varsigma}/2$. 
Without this step it would be impossible to find a simple transformation of the problem~\eqref{eq:dualaction} that generates the $M^{\varsigma}_{\nu\nu'\omega}$ correction in order to cancel the full local vertex function $\Gamma_{\nu\nu'\omega}$ from the theory.
This approximation is justified in the high-frequency limit where the fermion-boson vertex function $\Lambda_{\nu\omega}$  is equal to unity (Fig.~\ref{fig:Lambda}), and also by a good agreement of the resulting theory with much more elaborate approaches discussed below.
Following recent works~\cite{PhysRevLett.121.037204, PhysRevB.99.115124}, the $M^{\varsigma}_{\nu\nu'\omega}$ correction can be obtained with the help of an additional Hubbard--Stratonovich transformation over bosonic variables $\varphi^{\varsigma}\to{}b^{\varsigma}$ (for details, see Appendix~\ref{App:Action}). As a result, we get the final expression for the action of the effective fermion-boson model
\begin{align}
{\cal S}_{f\text{-}b} = &-\sum_{\kv,\nu,\sigma} f^{*}_{{\bf k}\nu\sigma}\tilde{\cal G}^{-1}_{\kv\nu\sigma}f^{\phantom{*}}_{\kv\nu\sigma} 
-\frac12\sum_{\qv,\omega,\varsigma} b^{\varsigma}_{\qv\omega}{\cal W}^{\varsigma ~ -1}_{\qv\omega}b^{\varsigma}_{-\qv,-\omega} \notag\\
&+\sum_{\kv,\qv}\sum_{\nu,\omega}\sum_{\varsigma,\sigma,\sigma'}
\Lambda^{\varsigma}_{\nu\omega} \, f^{*}_{\kv\nu\sigma}\,\sigma^{\varsigma}_{\sigma\sigma'} f^{\phantom{*}}_{\kv+\qv,\nu+\omega,\sigma'} \, b^{\varsigma}_{-\qv,-\omega}.
\label{eq:fbaction}
\end{align}
The bare Green's function $\tilde{\cal G}_{\kv\nu\sigma}$ remains unchanged during the last transformation, and the bare bosonic propagator becomes equal to ${\cal W}^{\varsigma}_{\qv\omega} = W^{\rm \varsigma\,EDMFT}_{\qv\omega} - U^{\varsigma}/2$. Note that if the local Coulomb interaction is considered in the Ising decoupling form, the bare bosonic propagator of the new fermion-boson theory coincides with the renormalized interaction of EDMFT ${\cal W}^{\varsigma}_{\qv\omega} = W^{\rm \varsigma\,EDMFT}_{\qv\omega}$ as discussed in Appendix~\ref{App:Action}.

The simplest set of diagrams for the self-energy and polarization operator has the following form
\begin{align}
\tilde{\Sigma}_{\kv\nu\sigma} &= -\sum_{\qv,\omega,\varsigma}\Lambda^{\varsigma}_{\nu+\omega,-\omega}\tilde{G}^{\phantom{2}}_{\kv+\qv,\nu+\omega,\sigma'}W^{\varsigma}_{\qv\omega}\Lambda^{\varsigma}_{\nu,\omega} 
= \vcenter{\hbox{\includegraphics[width=0.17\linewidth]{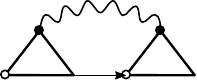}}} \notag\\
\tilde{\Pi}^{\varsigma}_{\qv\omega}  &= \sum_{\kv,\nu,\sigma(')} \Lambda^{\varsigma}_{\nu+\omega,-\omega}\tilde{G}^{\phantom{2}}_{\kv+\qv,\nu+\omega,\sigma}\tilde{G}^{\phantom{2}}_{\kv\nu\sigma'}\Lambda^{\varsigma}_{\nu,\omega} 
= \vcenter{\hbox{\includegraphics[width=0.17\linewidth]{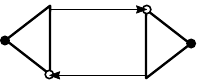}}} \label{eq:fbPi}
\end{align}
Here, $\tilde{G}_{\kv\nu\sigma}$ and $W^{\varsigma}_{\qv\omega}$ are full propagators of the derived fermion-boson problem~\eqref{eq:fbaction}. 
We prefer to keep fermions in the dual space, which results in the following connection between dual and lattice self energies $\Sigma^{\rm latt}_{\kv\nu\sigma} = \Sigma^{\rm imp}_{\nu} + \Sigma'_{\kv\nu\sigma}$, where $\Sigma'_{\kv\nu} = \tilde{\Sigma}_{\kv\nu} (1+g_{\nu}\tilde{\Sigma}_{\kv\nu})^{-1}$, as derived in Refs.~\onlinecite{Rubtsov20121320, PhysRevB.93.045107, PhysRevB.94.205110}. The last expression excludes the double counting between contributions of the local $\Sigma^{\rm imp}_{\nu}$ and nonlocal $\tilde{\Sigma}_{\kv\nu}$ self-energies to the lattice Green's function $G_{\kv\nu}$ that arise in the Dyson equation. Here, $g_{\nu}$ is the full local Green's function of the impurity problem. 
Although the introduced diagram for the nonlocal self-energy has a very simple form~\eqref{eq:fbPi}, it effectively contains the leading ``horizontal'' part of the two-particle ladder contribution that is present in much more advanced DF~\cite{PhysRevB.77.033101}, and DB~\cite{PhysRevB.72.035122, PhysRevLett.97.076405} theories (see Fig.~\ref{fig:GammaSigma}). 
Moreover, an account for this contribution does not require an inversion of the Bethe-Salpeter equation, which is a big advantage for numerical calculations. 

\begin{figure}[t!]
\includegraphics[width=0.9\linewidth]{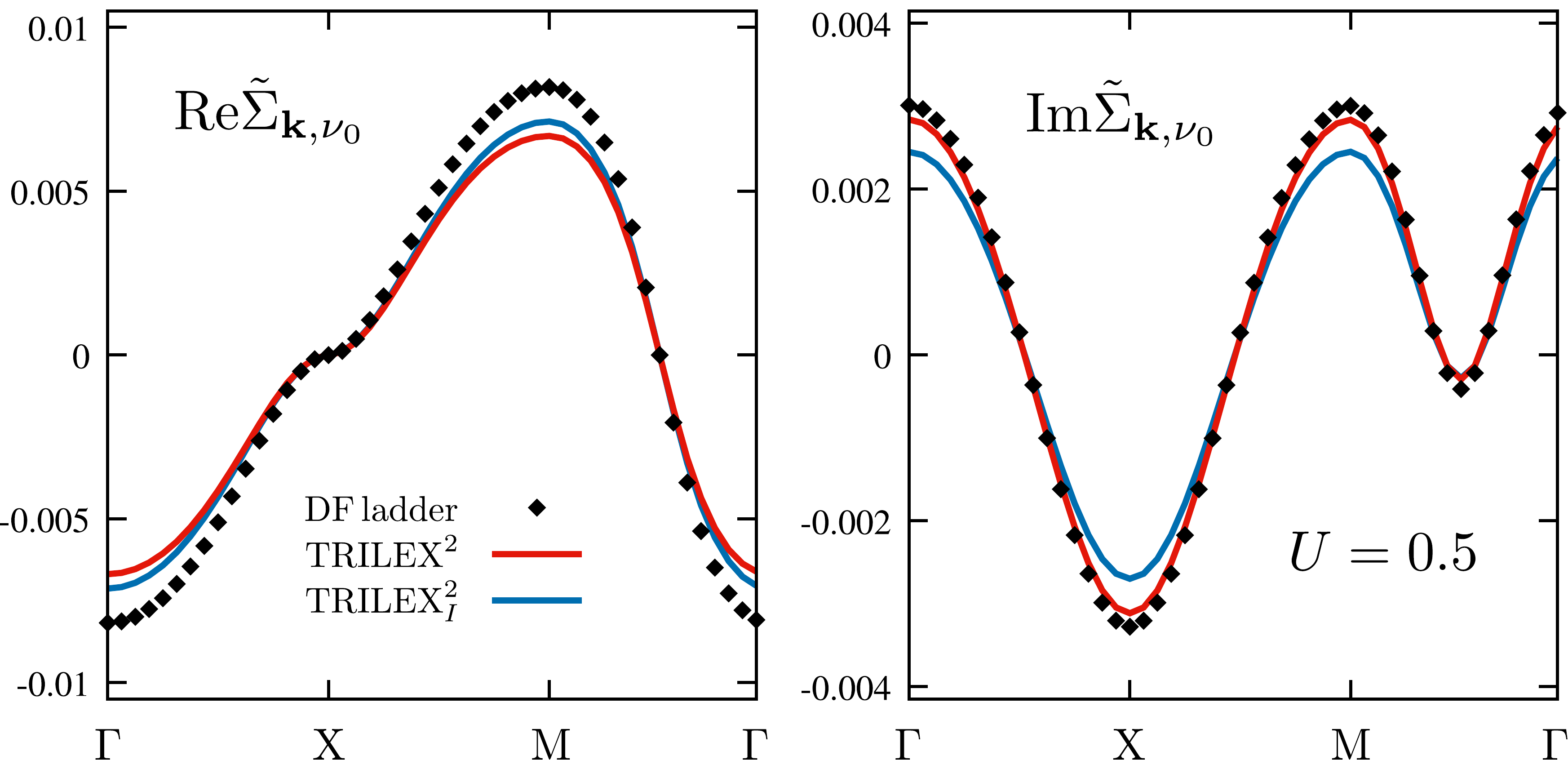} \\[0.3cm]
\includegraphics[width=0.9\linewidth]{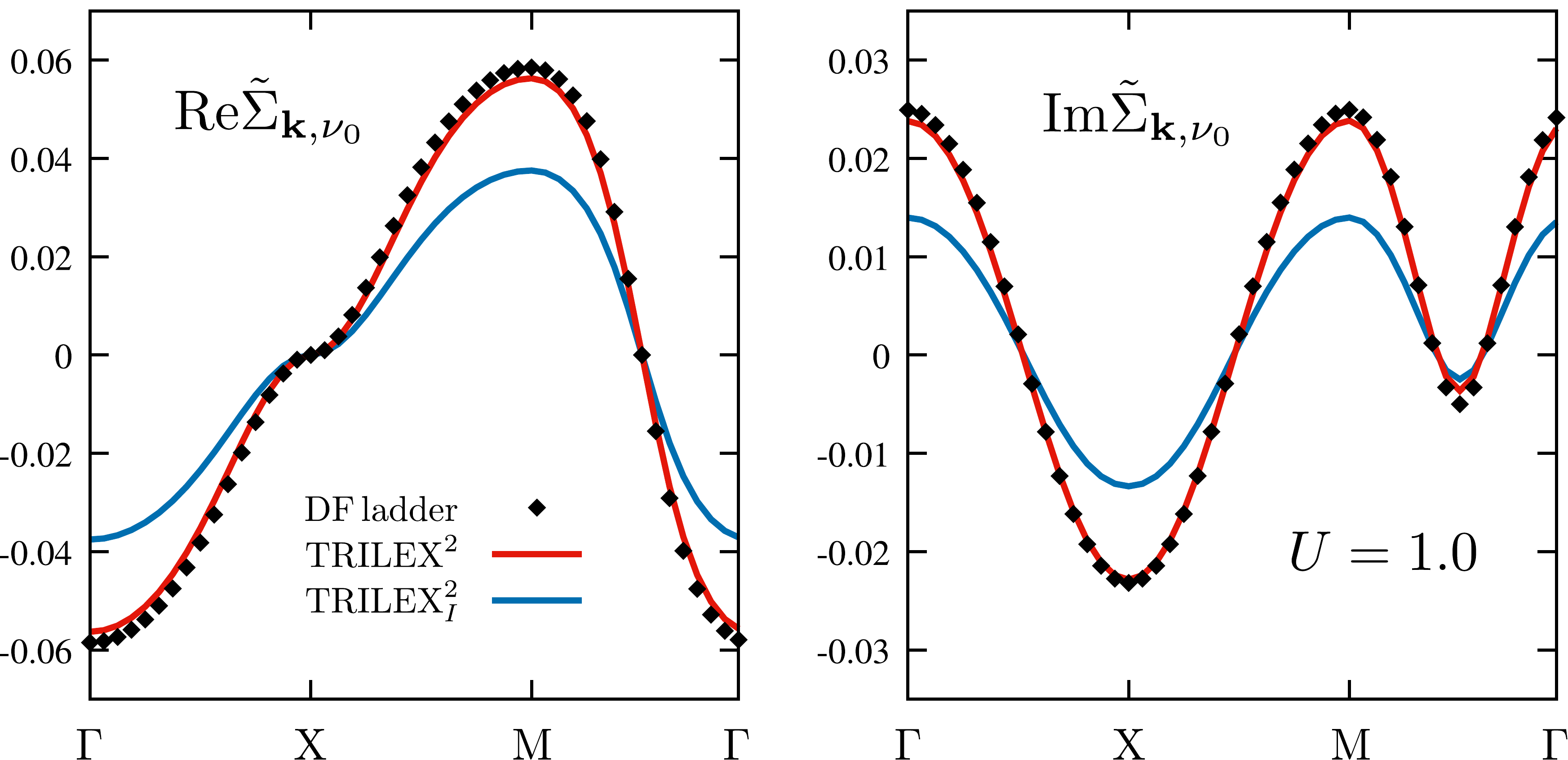} \\[0.3cm]
\includegraphics[width=0.9\linewidth]{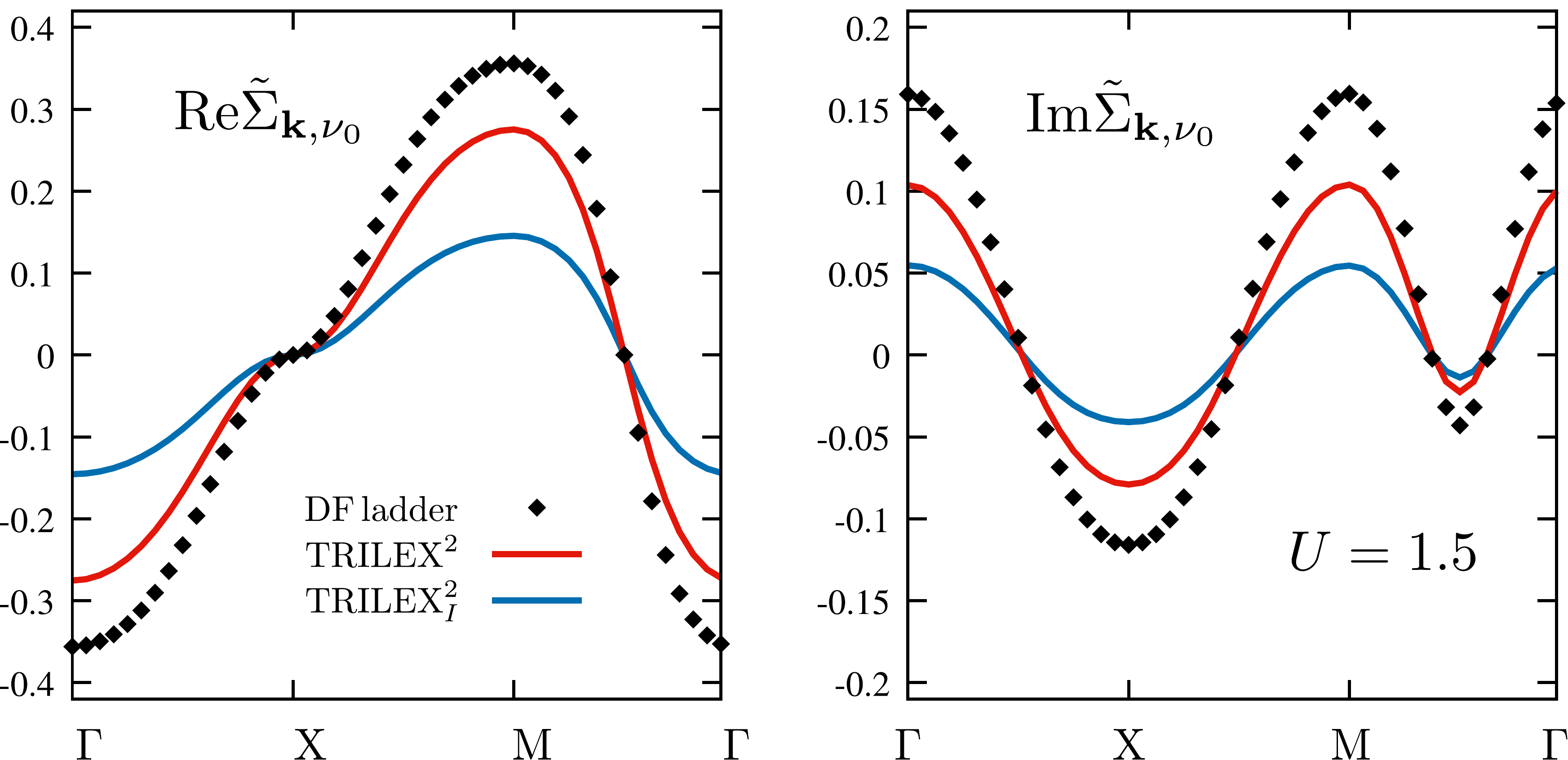}
\caption{Real and imaginary parts of the nonlocal self-energy $\tilde{\Sigma}_{\kv\nu}$ at the first fermionic Matsubara frequency $\nu_0$ obtained for the ladder dual fermion and TRILEX$^2$ (for the ``unique'' and Ising form of the bare interaction) approaches. Results are calculated for $U=0.5$ (top row), $U=1.0$ (middle row), and $U=1.5$ (bottom row). }
\label{fig:Sigma}
\end{figure}

At first glance, nonlocal diagrams introduced in Eq.~\eqref{eq:fbPi} do not obey the Hedin form~\cite{GW1}, where the full lattice fermion-boson vertex function appears only at one side of the diagram. However, in the resulting action~\eqref{eq:fbaction} the full local fermion-boson vertex $\Lambda^{\varsigma}_{\nu\omega}$ is the {\it bare} interaction vertex for an effective lattice problem that consequently enters diagrams for the self-energy and polarization operator from both sides. The importance to have the local vertex function at both sides of dual diagrams has been discussed in details in Ref.~\onlinecite{PhysRevB.94.205110}.

The present approach immediately suggests an improvement for already existing partially bosonized theories. Indeed, if two or one fermion-boson vertices in Eq.~\eqref{eq:fbPi} are replaced by unity, our method reduces to $GW$+DMFT or TRILEX approaches, respectively, but with a more accurate Fierz-ambiguity-free form of the bosonic propagator. Thus, we will call the introduced set of diagrams~\eqref{eq:fbPi} for the self-energy and polarization operator that contains a double triangular fermion-boson vertex correction as the TRILEX$^2$ approximation of the partially bosonized theory.

\section{Results}

\subsection{Nonlocal self-energy}
The performance of the TRILEX$^2$ approach can be tested against a more elaborate ladder DF method, which is accurate enough in the regime of strong interactions $U$ not exceeding the bandwidth ($U\leq2.0$)~\cite{PhysRevB.96.035152, PhysRevB.94.035102, PhysRevB.97.125114}. Figure~\ref{fig:Sigma} shows the nonlocal self-energy $\tilde{\Sigma}_{\kv\nu}$ at zero Matsubara frequency $\nu_0$ for different approaches. The result is obtained within a single-shot calculation performed on top of the converged DMFT solution, so that the local self-energy $\Sigma^{\rm imp}_{\nu}$ has the same value for all compared theories. For numerical solution of the impurity problem we used the open source CT-HYB solver~\cite{HAFERMANN20131280, PhysRevB.89.235128} based on ALPS libraries~\cite{Bauer_2011}. 

We find that the result for the self-energy of the TRILEX$^2$ approximation is in a very good agreement with the one of the ladder DF approach even in the strongly interacting $U=1.0$ regime. A small mismatch between these two results appears because the TRILEX$^2$ theory does not account for ``vertical'' contributions $M^{\varsigma}_{\nu,\nu+\omega,\nu'-\nu}$ to the fermion-fermion vertex that are present in the DF approach. The absence of these corrections only slightly modifies the result, but greatly simplifies numerical calculations. The effect of neglected contribution of vertical diagrams is more visible in the weakly-correlated regime ($U=0.5$). This can be explained by the fact that the horizontal contribution to the vertex function becomes leading when collective fluctuations in the corresponding channel are strong~\cite{PhysRevLett.121.037204, PhysRevB.99.115124}. In the regime of $U=0.5$ charge and spin fluctuations only start to develop, which results in a larger mismatch with the exact result for the self-energy. However, the value of the nonlocal self-energy in this regime is relatively small, so this inconsistency should not lead to a serious problem for calculation of physical observables. At larger value of the interaction ($U=1.5$) the contribution of vertical diagrams becomes more important. As expected, the TRILEX$^2_{I}$ result, which is based on the Ising decoupling, provides a less accurate result due to missing diagrams in $x$ and $y$ spin channels.

\subsection{Metal-to-Mott-insulator phase transition}
The present approach also shows a qualitatively good estimation for a metal to Mott insulator phase transition.
The corresponding phase boundary can be obtained from the behavior of the local Green's function at imaginary time $\tau=\beta/2$, which approximates the quasiparticle density of states at the Fermi level~\cite{PhysRevB.90.235135}. For this aim we perform a fully self-consistent TRILEX$^2$ calculation using a standard self-consistency condition on the local part of the lattice Green's function $\sum_{\kv}G_{\kv\nu}=g_{\nu}$ to determine the fermionic hybridization function $\Delta_{\nu}$ of the impurity problem. 
We find that in our case, the phase transition occurs at much smaller values of the local Coulomb interaction $U\simeq1.7$ compared to the DMFT result~\cite{PhysRevLett.101.186403}. The same trend and qualitatively similar results were previously reported for cluster DMFT~\cite{PhysRevLett.101.186403} and second-order DF~\cite{PhysRevB.98.155117} calculations. Surprisingly, the elimination of one fermion-boson vertex in diagrams~\eqref{eq:fbPi}, as originally proposed in the TRILEX approach~\cite{PhysRevB.92.115109, PhysRevB.93.235124, PhysRevLett.119.166401}, drastically changes the metal-to-Mott-insulator transition point and shifts it to a larger value of the local Coulomb interaction compared even to the DMFT result~\cite{PhysRevB.93.235124}. This can be attributed to the fact that the fermion-boson vertex at low frequencies considerably deviates from unity in the strongly interacting regime as shown in Fig.~\ref{fig:Lambda}.

\section{Conclusions}

To conclude, the derived fermion-boson theory is a powerful tool for description of many-body effects beyond the dynamical mean-field level. The main advantage is that the method does not suffer from the Fierz ambiguity problem, which is present in all partially bosonized theories. 
The TRILEX$^2$ approximation of the theory combines a simplicity of mean-field and $GW$-like diagrammatic descriptions of collective excitations with a high performance of the method comparable to much more elaborate approaches. 
A rigorous account for spin fluctuations in this approach provides an opportunity for a solution of a challenging problem of realistic magnetic $GW$-based calculations~\cite{PhysRevLett.100.116402, PhysRevB.95.041112}.
Finally, it is worth noting that the derived formalism is not restricted only to diagrams for the self-energy and polarization operator introduced in Eq.~\eqref{eq:fbPi}, respectively. The effective fermion-boson action~\eqref{eq:fbaction} also allows for a more advanced solution of the problem using, for example, functional renormalization group (fRG)~\cite{RevModPhys.84.299, doi:10.1080/00018732.2013.862020, PhysRevLett.112.196402, PhysRevB.91.045120, PhysRevB.99.115112, 10.21468/SciPostPhys.6.1.009}, parquet~\cite{PhysRevB.64.165107, PhysRevB.91.115115, PhysRevB.93.165103, PhysRevB.93.245102}, or diagrammatic Monte Carlo~\cite{PhysRevB.96.035152, PhysRevB.94.035102} methods.

\begin{acknowledgments}
The authors thank F. Krien and S. Brener for valuable comments on the work. The authors also thank M. Katsnelson and A. Rubtsov for inspiring discussions and long-term collaboration. The work of E.A.S. was supported by the Russian Science Foundation, Grant No. 17-72-20041. This work was partially supported by the Cluster of Excellence ``Advanced Imaging of Matter'' of the Deutsche Forschungsgemeinschaft (DFG) - EXC 2056 - Project No. ID390715994, and by North-German Supercomputing Alliance (HLRN) under the Project No. hhp00042.
\end{acknowledgments}

\appendix

\section{Approximation for the fermion-fermion vertex}
\label{App:Vertex}

In this Appendix we derive an approximation for the full local fermion-fermion vertex.
We start with the expression~\eqref{eq:BareGammaSpin} for the bare vertex $\Gamma^{\,0\,\varsigma}_{\nu\nu'\omega}$ of the impurity problem~\eqref{eq:action_imp_G}.
Using the exact relation between charge and spin components of the bare Coulomb interaction we find that the expression for this bare vertex does not depend on the performed decoupling of the local Coulomb interaction and contains the contribution of the full $U$ in all considered channels. 
This result is in agreement with the fact that the bare interaction in the Bethe-Salpeter equation for the susceptibility is given by the full local Coulomb interaction~\cite{PhysRevB.99.115124}.
In order to find the origin of the reducible contribution with respect to a bosonic line (hereinafter, we will call this contribution $w$-reducible) to the fermion-fermion vertex, let us dress the bare vertex in the corresponding ``horizontal'' particle-hole channel as  
\begin{align}
\overline{\Gamma}^{\,\varsigma}_{\nu\nu'\omega} = \sum_{\nu'',\nu'''} \Gamma^{0\,\varsigma}_{\nu\nu''\omega} \, \chi^{\,\varsigma}_{\nu''\nu'''\omega} \, \Gamma^{0\,\varsigma}_{\nu'''\nu'\omega},
\label{eq:Gscreened_app}
\end{align}
where
\begin{align}
\chi^{\,\varsigma}_{\nu\nu'\omega} = -\av{ \left(c^{*}_{\nu+\omega,\sigma_1}\sigma^{\varsigma}_{\sigma_1\sigma_2} c^{\phantom{*}}_{\nu,\sigma_2}\right) \, \left(c^{*}_{\nu'\sigma_3}\sigma^{\varsigma}_{\sigma_3\sigma_4} c^{\phantom{*}}_{\nu'+\omega,\sigma_4}\right)}_{\rm conn}
\label{eq:Xgen}
\end{align}
is a generalized susceptibility of the impurity problem in a corresponding channel. 
After the antisymmetrization, this screened vertex~\eqref{eq:Gscreened_app} together with the bare vertex $\Gamma^{\,0\,\varsigma}_{\nu\nu''\omega}$ makes up the simplest approximation for the fermion-fermion vertex function of the impurity problem
\begin{align}
\Gamma^{\,c}_{\nu\nu'\omega} &\simeq \Gamma^{0\,c}_{\nu\nu'\omega} + \frac12 \overline{\Gamma}^{\,c}_{\nu\nu'\omega} - \frac14 \overline{\Gamma}^{\,c}_{\nu,\nu+\omega,\nu'-\nu} - \frac34 \overline{\Gamma}^{\,s}_{\nu,\nu+\omega,\nu'-\nu},
\notag\\
\Gamma^{\,s}_{\nu\nu'\omega} &\simeq \Gamma^{0\,s}_{\nu\nu'\omega} + \frac12 \overline{\Gamma}^{\,s}_{\nu\nu'\omega} + \frac14 \overline{\Gamma}^{\,s}_{\nu,\nu+\omega,\nu'-\nu} - \frac14 \overline{\Gamma}^{\,c}_{\nu,\nu+\omega,\nu'-\nu}.
\label{eq:Gspapprox_app}
\end{align}
Since the bare vertex function does not depend on the decoupling, this approximation is valid for any decomposition of the local Coulomb interaction. In the absence of bosonic hybridizations $Y^{\varsigma}=0$, the bare fermion-fermion vertex can be simply replaced by the bare Coulomb interaction $\Gamma^{\,0\,c/s}_{\nu\nu'\omega} = \pm{}U$ as derived above. Then, the generalized susceptibility~\eqref{eq:Xgen} in the expression for the screened vertex~\eqref{eq:Gscreened_app} reduces to a bosonic susceptibility $\chi^{\,\varsigma}_{\omega}$, and the approximation for the full fermion-fermion vertex takes the following simple form:  
\begin{align}
\Gamma^{\,c}_{\nu\nu'\omega} &\simeq U + \frac12 U\chi^{c}_{\omega}U - \frac14 U\chi^{\,c}_{\nu'-\nu}U - \frac34 U\chi^{\,s}_{\nu'-\nu}U, \notag\\
\Gamma^{\, s}_{\nu\nu'\omega} &\simeq -U + \frac12 U\chi^{\,s}_{\omega}U + \frac14 U\chi^{\,s}_{\nu'-\nu}U - \frac14 U\chi^{\,c}_{\nu'-\nu}U.
\label{eq:screened_vertex_app}
\end{align}
This approximation fully coincides with the approximation obtained in the work~\cite{PhysRevLett.114.236402}. The only difference is that here we do not perform a bosonization of collective fluctuations in the particle-particle channel as discussed in the main text. We also note that the susceptibility defined in our work is two times larger than the one introduced in Ref.~\onlinecite{PhysRevLett.114.236402}. 

Importantly, in the framework of the fermion-boson theory the interaction is introduced as the bosonic propagator. Thus, bare charge and spin interactions that enter the bare fermion-fermion vertex $\Gamma^{\,0\,\varsigma}_{\nu\nu'\omega}$ have to be considered as ``horizontal'' ${\cal U}^{\varsigma}_{\omega}$ and ``vertical'' bosonic ${\cal U}^{\varsigma}_{\nu'-\nu}$ lines. In this case, a simple replacement of the bare fermion-fermion vertex by the full local Coulomb interaction is no longer possible. 
First, let us isolate the $w$-reducible contribution in the approximation for the fermion-fermion vertex~\eqref{eq:Gspapprox_app}. If we take only horizontal ($\omega$-dependent) terms ${\cal U}^{\varsigma}_{\omega}$ from the bare vertex $\Gamma^{\,0\,\varsigma}_{\nu\nu'\omega}$ in the expression~\eqref{eq:Gscreened_app}, the generalized susceptibility again reduces to the bosonic one, and the $w$-reducible part of the screened vertex~\eqref{eq:Gscreened_app} becomes   $\overline{\Gamma}^{\,\varsigma}_{\nu\nu'\omega} = 4 {\cal U}^{\varsigma}_{\omega} \, \chi^{\,\varsigma}_{\omega} \, {\cal U}^{\varsigma}_{\omega}$. Other $w$-reducible terms in the screened vertex~\eqref{eq:Gscreened_app} appear from $w$-reducible contributions to the generalized susceptibility $\chi^{\,\varsigma}_{\nu''\nu'''\omega}$. If the latter contains at least one horizontal bosonic line ${\cal U}^{\varsigma}_{\omega}$ on which it can be cut into two separate parts, the bare vertex $\Gamma^{\,0\,\varsigma}_{\nu\nu'\omega}$ in the expression~\eqref{eq:Gscreened_app} does not necessarily have to be $w$-reducible in order to make the total expression reducible with respect to a bosonic propagator. This leads to an additional fermion-boson vertex correction $\Lambda^{\varsigma}_{\nu\omega}$ to the previously derived approximation for the screened vertex
\begin{align}
\overline{\Gamma}^{\,\varsigma}_{\nu\nu'\omega} = 4 \Lambda^{\varsigma}_{\nu\omega} {\cal U}^{\varsigma}_{\omega} \, \chi^{\,\varsigma}_{\omega} \, {\cal U}^{\varsigma}_{\omega}\Lambda^{\varsigma}_{\nu'+\omega,-\omega} + 2 \Lambda^{\varsigma}_{\nu\omega} {\cal U}^{\varsigma}_{\omega}\Lambda^{\varsigma}_{\nu'+\omega,-\omega} - 2{\cal U}^{\varsigma}_{\omega}. 
\label{eq:Gscreenedvert_app}
\end{align}
The term $2{\cal U}^{\varsigma}_{\omega}$ is already contained in the bare vertex $\Gamma^{0\,\varsigma}_{\nu\nu'\omega}$ and introduced here to simplify the expression.
We note that Eq.~\eqref{eq:Gspapprox_app} is only an approximation for the exact charge and spin fermion-fermion vertex functions. The exact $w$-reducible contribution to the screened fermion-fermion vertex~\eqref{eq:Gscreened_app} is given by the expression
\begin{align}
\overline{\Gamma}^{\,\varsigma}_{\nu\nu'\omega} = 4 \Lambda^{\varsigma}_{\nu\omega} w^{\varsigma}_{\omega}\Lambda^{\varsigma}_{\nu'+\omega,-\omega} - 4{\cal U}^{\varsigma}_{\omega},
\label{eq:ExactGscreenedvert_app}
\end{align}
where $w^{\varsigma}_{\omega} = {\cal U}^{\varsigma}_{\omega} + {\cal U}^{\varsigma}_{\omega}\,\chi^{\varsigma}_{\omega} {\cal U}^{\varsigma}_{\omega}$ is the full renormalized interaction of the impurity problem, and $\Lambda^{\varsigma}_{\nu\omega}$ is the exact fermion-boson vertex of the problem. Here, the term $ 4{\cal U}^{\varsigma}_{\omega}$ is again excluded from the expression, since it is already contained in the (nonsymmetrized) bare interaction. 

The remaining part of the generalized susceptibility in the expression~\eqref{eq:Gspapprox_app} for the screened vertex is irreducible with respect to the bosonic propagator. Together with vertical lines ${\cal U}^{\varsigma}_{\nu'-\nu}$ from the bare fermion-fermion vertex $\Gamma^{0\,\varsigma}_{\nu\nu'\omega}$ it makes the $w$-irreducible contribution to the full fermion-fermion vertex function that is not accounted for by the fermion-boson theory. 
As discussed in the main text, the ladder-like irreducible contributions to the fermion-fermion vertex function can be fully excluded by a proper choice of the bare interaction $U^{c} = -U^{s} = U/2$ that has the same value for all ${s}=\{x, y, z\}$ spin components. Since this unique form of the bare interaction cannot be obtained by any of the decoupling of the local Coulomb interaction, we will make separate decouplings for every bosonic channel to keep the bare interaction in the proposed form. Then, coming back to a nonsymmetrized form of the bare fermion-fermion vertex function~\eqref{eq:BareGammaSpin}, we get $\Gamma^{\,0\,\varsigma}_{\nu\nu'\omega} = 2{\cal U}^{\varsigma}_{\omega} + 2Y^{\varsigma}_{\omega}$. Together with the screened interaction $\overline{\Gamma}^{\varsigma}_{\nu\nu'\omega}$ from~\eqref{eq:ExactGscreenedvert_app}, which is also written in the antisymmetrized form, it makes the total approximation for the nonsymmetrized full fermion-fermion vertex function
\begin{align}
\frac18\Gamma^{\,\varsigma}_{\nu\nu'\omega} \simeq  \frac12M^{\varsigma}_{\nu\nu'\omega} = \frac12\left(\Lambda^{\varsigma}_{\nu\omega} w^{\varsigma}_{\omega}\Lambda^{\varsigma}_{\nu'+\omega,-\omega} - U^{\varsigma}/2\right).
\label{eq:Gtot_app}
\end{align}
The term $U^{\varsigma}/2$ appears here, because we use separate mutually exclusive decouplings of the bare Coulomb interaction in different bosonic channels. This term avoids the double counting of the bare Coulomb interaction in the bare vertex $\Gamma^{\,0}_{\nu\nu'\omega}$.
Note that the same procedure can be performed for the Ising form of the bare interaction $U^{c} = -U^{z} = U/2$ and $U^{x}=U^{y}=0$. Since this form of decoupling is identical for all channels, this does not lead to a double counting of the local Coulomb interaction. Then, the approximation for the fermion-fermion vertex in the antisymmetrized form is given by the expression $M^{\varsigma}_{\nu\nu'\omega} = \Lambda^{\varsigma}_{\nu\omega} w^{\varsigma}_{\omega}\Lambda^{\varsigma}_{\nu'+\omega,-\omega}$.


The final expression for the $w$-reducible approximation of the full fermion-fermion vertex function can be obtained after antisymmetrizing the expression~\eqref{eq:Gtot_app}
\begin{align}
\Gamma^{\,c}_{\nu\nu'\omega} &= 2M^{c}_{\nu\nu'\omega} - M^{c}_{\nu,\nu+\omega,\nu'-\nu} - 3M^{s}_{\nu,\nu+\omega,\nu'-\nu}, \notag\\
\Gamma^{\,s}_{\nu\nu'\omega} &= 2M^{s}_{\nu\nu'\omega} + M^{s}_{\nu,\nu+\omega,\nu'-\nu} - M^{c}_{\nu,\nu+\omega,\nu'-\nu}. \label{eq:SpinGamma_app}
\end{align}
Note that  the $w$-reducible interaction~\eqref{eq:Gtot_app}, which is introduced to exclude the exact fermion-fermion vertex from the action, does not have a uniform structure due to a presence of the $- U^{\varsigma}/2$ term that does not contain fermion-boson vertex functions. Therefore, the correction $M^{\varsigma}_{\nu\nu'\omega}$ cannot be easily generated performing transformations of the lattice action discussed below. Thus, we make a small additional approximation for the $w$-reducible fermion-fermion vertex $M^{\varsigma}_{\nu\nu'\omega}\simeq\Lambda^{\varsigma}_{\nu\omega} \bar{w}^{\varsigma}_{\omega} \, \Lambda^{\varsigma}_{\nu'+\omega,-\omega}$ including the $U^{\varsigma}/2$ term in the propagator $\bar{w}^{\varsigma}_{\omega} = w^{\varsigma}_{\omega} - U^{\varsigma}/2$. After that, the exact~\eqref{eq:ExactGscreenedvert_app} expression for the reducible contribution to the fermion-fermion vertex function coincides with the approximate one derived in Eq.~\eqref{eq:Gscreenedvert_app}. In addition, the last approximation can be motivated by the asymptotic behavior of the fermion-boson vertex function $\Lambda_{\nu\omega}\to1$ at large frequencies. 

\onecolumngrid

\section{Derivation of the effective fermion-boson problem}
\label{App:Action}


In this Appendix we derive an effective fermion-boson problem. 
We start with two Hubbard-Stratonovich transformations of the nonlocal part of the lattice action of the extended Hubbard model~\eqref{eq:actionlatt}
\begin{align}
&\exp\left\{\sum_{\kv,\nu,\sigma} c^{*}_{\kv\nu\sigma}[\Delta^{\phantom{*}}_{\nu\sigma}-\varepsilon^{\phantom{*}}_{\kv}]c^{\phantom{*}}_{\kv\nu\sigma}\right\} = 
D_{f}\int D[f^{*},f] \exp\left\{-\sum_{\kv,\nu,\sigma}\left( f^{*}_{\kv\nu\sigma}g^{-1}_{\nu\sigma}[\Delta^{\phantom{*}}_{\nu\sigma}-\varepsilon^{\phantom{*}}_{\kv}]^{-1}g^{-1}_{\nu\sigma}f^{\phantom{*}}_{\kv\nu\sigma} + c^{*}_{\kv\nu\sigma}g^{-1}_{\nu\sigma}f^{\phantom{*}}_{\kv\nu\sigma} + f^{*}_{\kv\nu\sigma}g^{-1}_{\nu\sigma}c^{\phantom{*}}_{\kv\nu\sigma}\right)\right\}, \notag\\
&\exp\left\{\sum_{\qv,\omega,\varsigma}\frac12\,\rho^{\varsigma}_{\qv\omega}\left[Y^{\varsigma}_{\omega}-V^{\varsigma}_{\qv}\right]\rho^{\varsigma}_{-\qv,-\omega}\right\} = 
D_{\varphi}\int D[\phi^{\varsigma}] \exp\left\{-\sum_{\qv,\omega,\varsigma}\left( \frac12\,\varphi^{\,\varsigma}_{\qv\omega}\alpha^{\varsigma~-1}_{\omega}\left[Y^{\varsigma}_{\omega}-V^{\,\varsigma}_{\qv}\right]^{-1}\alpha^{\varsigma~-1}_{\omega}\varphi^{\,\varsigma}_{-\qv,-\omega} + 
\varphi^{\,\varsigma}_{\qv\omega}\alpha^{\varsigma~-1}_{\omega}\rho^{\varsigma}_{-\qv,-\omega}\right)\right\}, 
\end{align}
where terms $D_{f} = {\rm det}\left[g_{\nu}\left(\Delta_{\nu\sigma}-\varepsilon_{\kv}\right)g_{\nu}\right]$ and $D^{-1}_{\varphi} = \sqrt{{\rm det}\left[\alpha^{\varsigma}_{\omega}\left(Y^{\varsigma}_{\omega}-V^{\varsigma}_{\qv}\right)\alpha^{\varsigma}_{\omega}\right]}$ can be neglected when calculating expectation values. Here $g_{\nu}$ is the full local Green's function of the impurity problem. ${\cal U}^{\varsigma}_{\omega}=U^{\varsigma} +Y^{\varsigma}_{\omega}$, and $w^{\varsigma}_{\omega}$ are the bare and renormalized interactions of the local impurity interaction in the corresponding bosonic channel. Factors $g_{\nu}$ and $\alpha^{\varsigma}_{\omega} = w^{\varsigma}_{\omega}/{\cal U}^{\varsigma}_{\omega}$ in the Hubbard--Stratonovich transformations are introduced for the special reason to express the interaction part of the transformed action in terms of full local vertex function of the impurity problem~\cite{PhysRevB.94.205110}. After these transformations the action takes the following form
\begin{align} 
{\cal S}'
&= \sum_{i}{\cal S}^{(i)}_{\rm imp} +\sum_{\kv,\nu,\sigma} \left[ c^{*}_{\kv\nu\sigma} g^{-1}_{\nu\sigma}f^{\phantom{*}}_{\kv\nu\sigma} + f^{*}_{\kv\nu\sigma} g^{-1}_{\nu\sigma}c^{\phantom{*}}_{\kv\nu\sigma} \right] 
+\sum_{\qv,\omega,\varsigma}\varphi^{\varsigma}_{\qv\omega} \alpha^{\varsigma~-1}_{\omega}\rho^{\varsigma}_{-\qv,-\omega} \notag\\
&- \sum_{\kv,\nu,\sigma} f^{*}_{\kv\nu\sigma}g^{-1}_{\nu\sigma} [\varepsilon^{\phantom{*}}_{\kv}-\Delta^{\phantom{*}}_{\nu\sigma}]^{-1} g^{-1}_{\nu\sigma}f^{\phantom{*}}_{\kv\nu\sigma} 
-\frac12\sum_{\qv,\omega,\varsigma} \varphi^{\varsigma}_{\qv\omega}
\alpha_{\omega}^{\varsigma~-1} \left[V^{\varsigma}_{\qv} - Y^{\varsigma}_{\omega}\right]^{-1}\alpha_{\omega}^{\varsigma~-1}\varphi^{\varsigma}_{-\qv,-\omega}. 
\end{align}
The above introduced transformations allow to integrate out the impurity part of the problem as
\begin{align}
\int D[c^{*},c]\,&\exp\left\{-\sum_{i}{\cal S}^{(i)}_{\rm imp} - \sum_{\kv,\nu,\sigma} \left[ c^{*}_{\kv\nu\sigma} g^{-1}_{\nu\sigma}f^{\phantom{*}}_{\kv\nu\sigma} + 
f^{*}_{\kv\nu\sigma} g^{-1}_{\nu\sigma}c^{\phantom{*}}_{\kv\nu\sigma} \right] 
-\sum_{\qv,\omega,\varsigma} \varphi^{\varsigma}_{\qv\omega} \alpha_{\omega}^{\varsigma~-1} \rho^{\varsigma}_{-\qv,-\omega} \right\} = \notag\\
{\cal Z}_{\rm imp} \times &\exp\left\{ -\sum_{\kv,\nu,\sigma}f^{*}_{\kv\nu\sigma} g^{-1}_{\nu\sigma}f^{\phantom{*}}_{\kv\nu\sigma} 
-\frac12\sum_{\qv,\omega,\varsigma} \varphi^{\varsigma}_{\qv\omega} \alpha_{\omega}^{\varsigma~-1} \chi^{\,\varsigma}_{\omega}\alpha_{\omega}^{\varsigma~-1} \varphi^{\varsigma}_{-\qv,-\omega} - \tilde{\cal F}[f,\varphi]\right\},
\end{align}
where ${\cal Z}_{\rm imp}$ is a partition function of the impurity problem.
Here, the interaction part of the action $\tilde{\cal F}[f,\varphi]$ contains an infinite series of full vertex functions of impurity problem as discussed in~\cite{Rubtsov20121320, PhysRevB.93.045107}. The lowest order interaction terms are
\begin{align}
\tilde{\cal F}[f,\varphi]
&\simeq\sum_{\kv,\kv',\qv}\sum_{\nu,\nu',\omega}\sum_{\varsigma,\sigma(')}
\left(\Lambda^{\varsigma}_{\nu\omega}\, f^{*}_{\kv\nu\sigma}f^{\phantom{*}}_{\kv+\qv,\nu+\omega,\sigma'}\,\varphi^{\varsigma}_{-\qv,-\omega} 
+ \frac14 \Gamma^{\,\sigma\sigma'\sigma''\sigma'''}_{\nu\nu'\omega} f^{*}_{\kv\nu\sigma}f^{\phantom{*}}_{\kv+\qv,\nu+\omega,\sigma'}f^{*}_{\kv'+\qv,\nu'+\omega,\sigma''} f^{\phantom{*}}_{\kv'\nu'\sigma'''}\right),
\label{eq:lowestint}
\end{align}
where the fermion-fermion and fermion-boson vertices have the following form
\begin{align}
\Gamma_{\nu\nu'\omega} = \frac{\av{c^{\phantom{*}}_{\nu\sigma} c^{*}_{\nu+\omega,\sigma'} c^{*}_{\nu'\sigma'''}c^{\phantom{*}}_{\nu'+\omega,\sigma''}}_{\rm c~ imp}}{g_{\nu\sigma}g_{\nu+\omega,\sigma'}g_{\nu'+\omega,\sigma''} g_{\nu'\sigma'''}},~~~
\Lambda^{\varsigma}_{\nu\omega} =  
\frac{\av{c^{\phantom{*}}_{\nu\sigma} c^{*}_{\nu+\omega,\sigma'}\,\rho^{\varsigma}_{\omega}}_{\rm imp}}{g_{\nu\sigma}\,g_{\nu+\omega,\sigma'}\alpha_{\omega}^{\varsigma}}.
\end{align}
%
Then, the initial lattice problem transforms to the following {\it dual} action 
\begin{align}
\label{eq:dualaction_app}
{\cal \tilde{S}}
= -\sum_{\kv,\nu,\sigma} f^{*}_{\kv\nu\sigma}\tilde{\cal G}^{-1}_{\kv\nu\sigma}f^{\phantom{*}}_{\kv\nu\sigma} 
-\frac12\sum_{\qv,\omega,\varsigma}  \varphi^{\varsigma}_{\qv\omega}
\tilde{\cal W}^{\varsigma~-1}_{\qv\omega}
 \varphi^{\varsigma}_{-\qv,-\omega} + \tilde{\cal F}[f,\varphi].
\end{align}
Here, bare propagators $\tilde{\cal G}_{\kv\nu\sigma} = G^{\rm EDMFT}_{\kv\nu\sigma} - g^{\phantom{E}}_{\nu\omega}$ and $\tilde{\cal W}^{\varsigma}_{\qv\omega} = W^{\varsigma \, \rm EDMFT}_{\qv\omega} - w^{\varsigma}_{\omega}$ are nonlocal parts of the Green's function $G^{\rm EDMFT}_{\kv\nu\sigma}$ and renormalized interaction $W^{\varsigma \, \rm EDMFT}_{\qv\omega}$ of EDMFT defined as
\begin{align}
G^{\rm EDMFT~-1}_{\kv\nu\sigma} = i\nu+\mu-\varepsilon_{\kv}-\Sigma^{\rm imp}_{\nu\sigma}, ~~~
W^{\varsigma \, \rm EDMFT~-1}_{\qv\omega} = \left(U^{\varsigma} + V^{\varsigma}_{\qv}\right)^{-1}-\Pi^{\varsigma \, \rm imp}_{\omega}.
\label{eq:EDMFTGW_app}
\end{align}
Here, $g_{\nu}$ and $w^{\varsigma}_{\omega}$ are the full local impurity Green's function and renormalized interaction of the impurity problem
\begin{align}
g^{-1}_{\nu\sigma} = i\nu+\mu-\Delta_{\nu}-\Sigma^{\rm imp}_{\nu\sigma}, ~~~
w^{\varsigma~-1}_{\omega} = \left(U^{\varsigma} + Y^{\varsigma}_{\omega}\right)^{-1}-\Pi^{\varsigma \, \rm imp}_{\omega}.
\label{eq:ImpGW_app}
\end{align}

The second transformation of bosonic variables that excludes the fermion-fermion vertex function from the dual action can be performed as follows. Let us add and subtract the term $\frac12\sum_{\qv,\omega,\varsigma} \varphi^{\varsigma}_{\qv\omega} \bar{w}_{\omega}^{\varsigma~-1}\varphi^{\varsigma}_{-\qv,-\omega}$ in the dual action
\begin{align} 
{\cal \tilde{S}}
&= -\sum_{\kv,\nu,\sigma} f^{*}_{\kv\nu\sigma}\tilde{\cal G}^{-1}_{\kv\nu\sigma}f^{\phantom{*}}_{\kv\nu\sigma}  
+\frac12\sum_{\qv,\omega,\varsigma} \varphi^{\varsigma}_{\qv\omega} \bar{w}_{\omega}^{\varsigma~-1}\varphi^{\varsigma}_{-\qv,-\omega}
+ \tilde{\cal F}[f,\varphi] 
-\frac12\sum_{\qv,\omega,\varsigma} \varphi^{\varsigma}_{\qv\omega}
\alpha^{\varsigma~-1}_{\omega} \left\{\left[V^{\varsigma}_{\qv} - Y^{\varsigma}_{\omega}\right]^{-1} - \chi^{\,\varsigma}_{\omega} + \alpha^{\varsigma}_{\omega}\bar{w}_{\omega}^{\varsigma~-1}\alpha^{\varsigma}_{\omega} \right\} \alpha_{\omega}^{\varsigma~-1} \varphi^{\varsigma}_{-\qv,-\omega}
\end{align}
Then, we can perform the following Hubbard-Stratonovich transformation
\begin{align}
&\exp\left\{\frac12\sum_{\qv,\omega,\varsigma} \varphi^{\varsigma}_{\qv\omega}
\alpha^{\varsigma~-1}_{\omega} \left\{\left[V^{\varsigma}_{\qv} - Y^{\varsigma}_{\omega}\right]^{-1} - \chi^{\,\varsigma}_{\omega} + \alpha^{\varsigma}_{\omega}\bar{w}_{\omega}^{\varsigma~-1}\alpha^{\varsigma}_{\omega} \right\} \alpha_{\omega}^{\varsigma~-1} \varphi^{\varsigma}_{-\qv,-\omega}\right\} = \notag\\
&D_{b}
\int D[b^{\varsigma}]\exp\left\{-\sum_{\qv,\omega,\varsigma}\left(\frac12\, b^{\varsigma}_{\qv\omega}\bar{w}^{-1}_{\omega}
\alpha^{\varsigma}_{\omega} \left\{\left[V^{\varsigma}_{\qv} - Y^{\varsigma}_{\omega}\right]^{-1} - \chi^{\,\varsigma}_{\omega} + \alpha^{\varsigma}_{\omega}\bar{w}_{\omega}^{\varsigma~-1}\alpha^{\varsigma}_{\omega} \right\}^{-1} \alpha_{\omega}^{\varsigma} \bar{w}^{-1}_{\omega}b^{\varsigma}_{-\qv,-\omega} 
- \varphi^{\varsigma}_{\qv\omega} \bar{w}^{-1}_{\omega} b^{\varsigma}_{-\qv,-\omega}\right)\right\}. 
\end{align}
The action transforms to
\begin{align} 
{\cal \tilde{S}}'
= &-\sum_{\kv,\nu,\sigma} f^{*}_{\kv\nu\sigma}\tilde{\cal G}^{-1}_{\kv\nu\sigma}f^{\phantom{*}}_{\kv\nu\sigma}  
+\frac12\sum_{\qv,\omega,\varsigma} b^{\varsigma}_{\qv\omega}\bar{w}^{-1}_{\omega}
\alpha^{\varsigma}_{\omega} \left\{\left[V^{\varsigma}_{\qv} - Y^{\varsigma}_{\omega}\right]^{-1} - \chi^{\,\varsigma}_{\omega} + \alpha^{\varsigma}_{\omega}\bar{w}_{\omega}^{\varsigma~-1}\alpha^{\varsigma}_{\omega} \right\}^{-1} \alpha_{\omega}^{\varsigma} \bar{w}^{-1}_{\omega}b^{\varsigma}_{-\qv,-\omega} \notag\\
&+\frac12\sum_{\qv,\omega,\varsigma} \varphi^{\varsigma}_{\qv\omega} \bar{w}_{\omega}^{\varsigma~-1}\varphi^{\varsigma}_{-\qv,-\omega}
- \sum_{\qv,\omega,\varsigma}\varphi^{\varsigma}_{\qv\omega} \bar{w}^{-1}_{\omega} b^{\varsigma}_{-\qv,-\omega} + \tilde{\cal F}[f,\varphi] 
\label{eq:prelastS_app}
\end{align}

Finally, bosonic fields $\varphi^{\varsigma}$ can be integrated out with respect to the Gaussian bosonic part of the dual action as
\begin{align}
&\int D[\varphi^{\varsigma}]\,\exp\left\{-\frac12\sum_{\qv,\omega,\varsigma} \varphi^{\varsigma}_{\qv\omega} \bar{w}_{\omega}^{\varsigma~-1}\varphi^{\varsigma}_{-\qv,-\omega} + \sum_{\qv,\omega,\varsigma}\varphi^{\varsigma}_{\qv\omega} \bar{w}^{-1}_{\omega} b^{\varsigma}_{-\qv,-\omega}  - \tilde{\cal F}[f,\varphi]  
\right\}
= 
{\cal Z}_{\varphi} \times \exp\left\{\frac12\sum_{\qv,\omega,\varsigma} b^{\varsigma}_{\qv\omega} \bar{w}_{\omega}^{\varsigma~-1}b^{\varsigma}_{-\qv,-\omega} - {\cal F}[f,b] \right\},
\label{eq:integrationphi}
\end{align}
where ${\cal Z}_{\varphi}$ is a partition function of the Gaussian part of the bosonic action. The integration of dual bosonic fields modifies the interaction that now has the following form
\begin{align}
{\cal F}[f,b]
&=\sum_{\kv,\qv}\sum_{\nu,\omega}\sum_{\varsigma,\sigma,\sigma'}
\Lambda^{\varsigma}_{\nu\omega} \, f^{*}_{\kv\nu\sigma}\sigma^{\varsigma}_{\sigma\sigma'} f^{\phantom{*}}_{\kv+\qv,\nu+\omega,\sigma'} \, b^{\varsigma}_{-\qv,-\omega} \notag\\
&+ \frac18 \sum_{\kv,\kv',\qv} \sum_{\nu,\nu',\omega} \sum_{\varsigma,\sigma(')} \left(  \Gamma^{\,\varsigma}_{\nu\nu'\omega} - 4M^{\varsigma}_{\nu\nu'\omega} \right) f^{*}_{\kv\nu\sigma}\sigma^{\varsigma}_{\sigma\sigma'}f^{\phantom{*}}_{\kv+\qv,\nu+\omega,\sigma'}\,f^{*}_{\kv'+\qv,\nu'+\omega,\sigma''} \sigma^{\varsigma}_{\sigma''\sigma'''} f^{\phantom{*}}_{\kv'\nu'\sigma'''}.
\label{eq:Wfull}
\end{align}
The $4M^{\varsigma}_{\nu\nu'\omega}$ term that was introduced in~\eqref{eq:Gtot_app} is exactly the approximation that excludes the full fermion-fermion vertex $\Gamma^{\,\varsigma_{\nu\nu'\omega}}$.
After collecting and simplifying all terms, the action~\eqref{eq:prelastS_app} takes a very compact form
\begin{align}
{\cal S}_{f\text{-}b} = &-\sum_{\kv,\nu,\sigma} f^{*}_{{\bf k}\nu\sigma}\tilde{\cal G}^{-1}_{\kv\nu\sigma}f^{\phantom{*}}_{\kv\nu\sigma} 
-\frac12\sum_{\qv,\omega,\varsigma} b^{\varsigma}_{\qv\omega}{\cal W}^{\varsigma ~ -1}_{\qv\omega}b^{\varsigma}_{-\qv,-\omega}
+\sum_{\kv,\qv}\sum_{\nu,\omega}\sum_{\varsigma,\sigma,\sigma'}
\Lambda^{\varsigma}_{\nu\omega}\, f^{*}_{\kv\nu\sigma}\sigma^{\varsigma}_{\sigma\sigma'} f^{\phantom{*}}_{\kv+\qv,\nu+\omega,\sigma'} \, b^{\varsigma}_{-\qv,-\omega} ,
\label{eq:fbaction_app}
\end{align}
where the bare bosonic propagator is equal to ${\cal W}^{\varsigma}_{\qv\omega} = \tilde{\cal W}^{\varsigma}_{\qv\omega} + \bar{w}_{\omega}^{\varsigma}$, which can also be rewritten as ${\cal W}^{\varsigma}_{\qv\omega} = W^{\varsigma \, \rm EDMFT}_{\qv\omega} - U^{\varsigma}/2$ for our choice $U^{c/s}=\pm{}U/2$ of the bare interaction. Since for the Ising decoupling $\bar{w}_{\omega}^{\varsigma}=w_{\omega}^{\varsigma}$, the bare bosonic propagator coincides with the renormalized interaction of EDMFT ${\cal W}^{\varsigma}_{\qv\omega} = W^{\varsigma \, \rm EDMFT}_{\qv\omega}$.

Remarkably, for our unique choice of the bare interaction $U^{\varsigma}$ the renormalized interaction of EDMFT 
can be identically rewritten in the form using in FLEX approach~\cite{BICKERS1989206, PhysRevB.57.6884}
\begin{align}
W^{\rm \varsigma\,EDMFT}_{\qv\omega} = \frac12  \hat{U}^{\varsigma}_{\qv}\left[1 - \hat{\Pi}^{\varsigma\,\rm imp}_{\omega} \hat{U}^{\varsigma}_{\qv}\right]^{-1},
\end{align}
where $\hat{U}^{c/s}_{\qv}=\pm{}U + 2V^{c/s}_{\qv}$ and $\hat{\Pi}^{\varsigma\,\rm imp}_{\omega}=\Pi^{\varsigma\,\rm imp}_{\omega}/2$ are the bare interaction and local polarization operator in FLEX notations. Thus, the introduced theory can be seen as an efficient combination of FLEX approach for local degrees of freedom with $GW$-like description of nonlocal fluctuations beyond the EDMFT level and additionally accounts for the fermion-boson vertex corrections.


\twocolumngrid

\bibliography{Charge_GW}

\end{document}